\documentclass[letterpaper, 10pt, conference]{IEEEtran}
\IEEEoverridecommandlockouts
\usepackage[centertags]{amsmath}
\usepackage{amssymb,amsthm,times}
\usepackage{enumitem}
\usepackage{float}
\usepackage{enumerate}
\usepackage{graphicx,subfigure}
\usepackage{nomencl} 
\usepackage{siunitx}
\usepackage{array}
\usepackage{tikz}
\usepackage{import}
\usepackage[hidelinks]{hyperref}
\usepackage{cleveref}
\allowdisplaybreaks
\usetikzlibrary{shapes,arrows,decorations.markings}
\graphicspath{{./Figs/}}

\title{\LARGE \bf
    Dynamics and Control of a Flapping Wing UAV with Abdomen Undulation Inspired by Monarch Butterfly
}

\author{Tejaswi K. C., Chang-kwon Kang, and Taeyoung Lee
    \thanks{Tejaswi K. C. and Taeyoung Lee, Mechanical and Aerospace Engineering, The George Washington University, Washington DC 20052 {\tt kctejaswi999@gmail.com,tylee@gwu.edu}}%
    \thanks{Chang-kwon Kang, Mechanical and Aerospace Engineering, University of Alabama in Huntsville, Huntsville AL 35899 {\tt ck0025@uah.edu}}%
    \thanks{\textsuperscript{\footnotesize\ensuremath{*}}This research has been supported in part by NSF under the grants NSF CMMI-1761618 and CMMI-1760928.}
}

\newcommand{\scalefont}[1]{\scalebox{1.5}{#1}}

\newcommand{\norm}[1]{\ensuremath{\left\| #1 \right\|}}

\newcommand{\braces}[1]{\ensuremath{\left\{ #1 \right\}}}

\newcommand{\pair}[1]{\ensuremath{\langle #1 \rangle}}

\newcommand{\SO}{\ensuremath{\mathsf{SO(3)}}}
\newcommand{\T}{\ensuremath{\mathsf{T}}}
\renewcommand{\L}{\ensuremath{\mathsf{L}}}
\newcommand{\so}{\ensuremath{\mathfrak{so}(3)}}

\renewcommand{\Re}{\ensuremath{\mathbb{R}}}

\newcommand{\D}{\ensuremath{\mathbf{D}}}

\newcommand{\ad}{\ensuremath{\mathrm{ad}}}

\newcommand{\G}{\ensuremath{\mathsf{G}}}
\newcommand{\g}{\ensuremath{\mathfrak{g}}}

\begin{document}

\maketitle

\begin{abstract}
    This paper presents a dynamic model and a control system for a flapping-wing unmanned aerial vehicle. 
    Inspired by flight characteristics captured from live Monarch butterflies, a new dynamic model is presented to account the effects of low-frequency flapping and abdomen undulation. 
    We developed it according to Lagrangian mechanics on a Lie group to obtain an elegant, global formulation of dynamics. 
    Then, a feedback control system is presented to asymptotically stabilize periodic motions with active motion of abdomen, and its stability is verified according to Floquet theory. 
    In particular, it is illustrated that the abdomen undulation has the desirable effects of reducing the variation of the total energy and also improving the stability of the proposed control system. 
\end{abstract}

\section{Introduction}

Flight controls of flapping wing aerial vehicles are challenging as they are essentially infinite dimensional, nonlinear time-varying systems, where the equations of motion describing displacement and the deformation of a flexible multi-body system are coupled with the Navier-Stokes equations.
As such, stability analyses of such FWUAVs rely on various assumptions~\cite{sun2007dynamic}.
Most of the current flight dynamics and control of FWUAVs have been conducted by linearizing the dynamic model around a selected operating point, and taking the average over a cycle of flapping \cite{taha2012flight,sun2014insect,doman2010wingbeat}.
For instance, a longitudinal flight control has been designed using the time-averaging theory~\cite{khan2007control}.

Recent works include adaptive controller implementation by employing neural networks along with disturbance observers~\cite{he2017adaptive}, and a path tracking control based on learning~\cite{lee2018learning}.
These work exploit the large disparity in time scales of wingbeat frequency and flight dynamics by utilizing high frequency oscillations of small wings.

On the other hand, there have been a few studies for the interaction of abdomen with the remaining body and wings. 
It has been shown that abdomen undulation may reduce power consumption from the dynamic coupling of wing-body motion~\cite{sridhar2019beneficial}.
Also, a simple two-dimensional model has been utilized to understand pitch instability of thorax and the effects of abdominal controls~\cite{jayakumar2018control}.
It is further reported that moths actively modulate their body shape to control flight in response to visual pitch stimuli, and it may contribute to pitch stability~\cite{dyhr2013flexible}.

In this paper, we present a dynamic model and a control system inspired by Monarch butterflies. 
They exhibit remarkable flight performances, migrating over a long range (up to 4000\;km) at very high altitudes while increasing aerodynamic efficiency.
The flight of Monarch butterflies is characterized by low flapping frequencies (10\;Hz), relatively large wingspans, and active abdomen undulation.
And as such, the existing approaches relying on the linearized dynamics over a short flapping period
are not suitable.

We first model a FWUAV as an articulated rigid body composed of the head/thorax, abdomen, and two wings that are interconnected by spherical joints, where the material parameters, size, and shape are selected to resemble those of Monarch butterflies.
Then, its dynamics are studied using quasi-steady aerodynamics assumptions and  Lagrangian mechanics on a manifold~\cite{sridhar2020geometric}.
This avoids complexities and singularities associated with local coordinates in representing complex maneuvers involving the dynamic coupling effects of abdomen and low-frequency flapping. 
Also, it provides a computationally efficient model that is suitable for design and verification of nonlinear feedback control laws.

Next, we design a nonlinear control system for the presented articulated rigid body model. 
In contrast to the current systems based-on on linearized dynamics, the proposed feedback control system utilizes Floquet theory to ensure stability of the controlled periodic orbit.
More specifically, the control input is formulated as the set of torques acting at each joint, and as such, it yields an optimal motion of the thorax and abdomen integrated with the wing flapping, thereby resembling the  distinct flight characteristics of Monarch.

Furthermore, we carefully analyze the effects of abdomen undulation in the periodic motion and the stability.
We show that abdomen undulation improves an energy efficiency of flight reducing the variation of the total energy and the power over a flapping period, and it further improves stability properties by enhancing the rate of convergence and by enlarging the region of attraction. 
Such advantageous effects of abdomen undulation in the controlled dynamics of FWUAVs have not been reported before. 

\section{Dynamics of Flapping-Wing UAV}\label{sec:dynamics}

Here, we present an articulated rigid body model for a flapping wing aerial vehicle~\cite{sridhar2020geometric}.
Throughout this paper, the three-dimensional special orthogonal group is denoted by $ \SO = \lbrace R \in \mathbb{R}^{3\times3} \mid R^T R = I, \det(R) = 1 \rbrace $, 
and the corresponding Lie algebra is $ \mathfrak{so}(3) = \lbrace A \in \mathbb{R}^{3\times3} \mid A = -A^T \rbrace $.
The \textit{Hat} map $ \wedge : \mathbb{R}^3 \to \mathfrak{s0}(3) $ is defined such that $\hat x y = x\times y$ for any $x,y\in\Re^3$.
And its inverse map is the \textit{vee} map, $ \vee:\so\rightarrow\Re^3 $.
Next, $e_i\in\Re^n$ denotes the $i$-th standard basis of $\Re^n$ for an appropriate dimension $n$, e.g., $e_1=(1,0,\ldots, 0)\in\Re^n$. 
Throughout this paper, the units are in $\si{kg}$, $\si{m}$, $\si{s}$, and $\si{rad}$, unless specified otherwise.

\subsection{Multibody Formulation}
Consider a flapping wing UAV that is composed of a body, an abdomen, and two wings attached to the body. 
Here, we assume that the head and thorax are combined into a single rigid body referred to as the \textit{body}. 

Define an inertial frame $\mathcal{F}_I=\{\mathbf{i}_x,\mathbf{i}_y,\mathbf{i}_z\}$, which is compatible to the NED (north-east-down) frame. The various components of this model are described below.

\begin{itemize}
	\item \textit{Body:}
	The origin of the body-fixed frame $\mathcal{F}_B=\{\mathbf{b}_x,\mathbf{b}_y,\mathbf{b}_z\}$is defined at the mass center of the body.
	Its attitude is given by $R\in\SO$ and the position of mass center is given by $ x \in \Re^3 $ in $ \mathcal{F}_I $. 
	The kinematics of the attitude is 
	\begin{align}
	\dot R = R \hat \Omega,
	\end{align}
    where $\Omega\in\Re^3$ is the angular velocity of the body resolved in $\mathcal{F}_B$.

	\item \textit{Right wing:}
	Let $\mathcal{F}_R=\{\mathbf{r}_x,\mathbf{r}_y,\mathbf{r}_z\}$ be the frame fixed to the right wing at its root. 
    And let $\mathcal{F}_S=\{\mathbf{s}_x,\mathbf{s}_y,\mathbf{s}_z\}$ be the stroke frame obtained by translating the origin of $\mathcal{F}_B$ to the center of the left and the right wing roots, and rotating it about $\mathbf{b}_y$ by a fixed angle $\beta\in[-\pi,\pi)$.
    Let $\mu_R\in\Re^3$ be the fixed vector from the origin of $\mathcal{F}_B$ to that of $\mathcal{F}_R$.

	\setlength{\unitlength}{0.1\textwidth}
	\begin{figure}
		\centerline{
			\footnotesize
			\subfigure[flapping angle,\newline $\phi_R\in[-\pi,\pi)$]{
			\scalebox{0.5}{
				\begin{picture}(3,2.6)(0,0)
				\put(0,0){\includegraphics[trim={4cm 3cm 4cm 1.5cm},clip,width=0.3\textwidth]{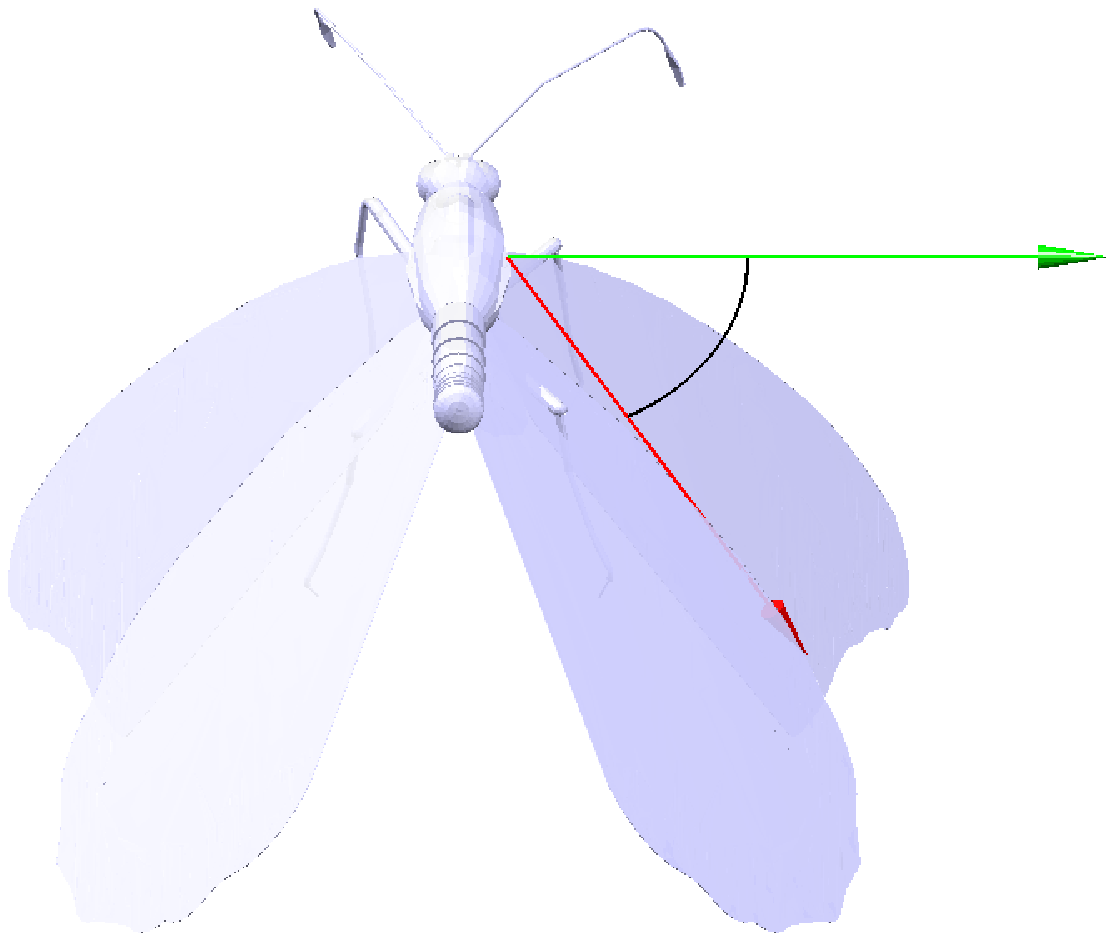}}
				\put(1.9,1.4){\scalefont{$\phi>0$}}
				\put(2.75,1.9){\scalefont{$\mathbf{s}_y$}}
				\put(2.2,0.6){\scalefont{$\mathbf{r}_y$}}
				\end{picture}
			}
			}
			\subfigure[pitch angle,\newline $\theta_R\in[-\pi,\pi)$]{
			\scalebox{0.5}{
		        \begin{picture}(3,2.6)(0,0)
				\put(0,0){\includegraphics[trim={4cm 3cm 2.5cm 1.5cm},clip,width=0.3\textwidth]{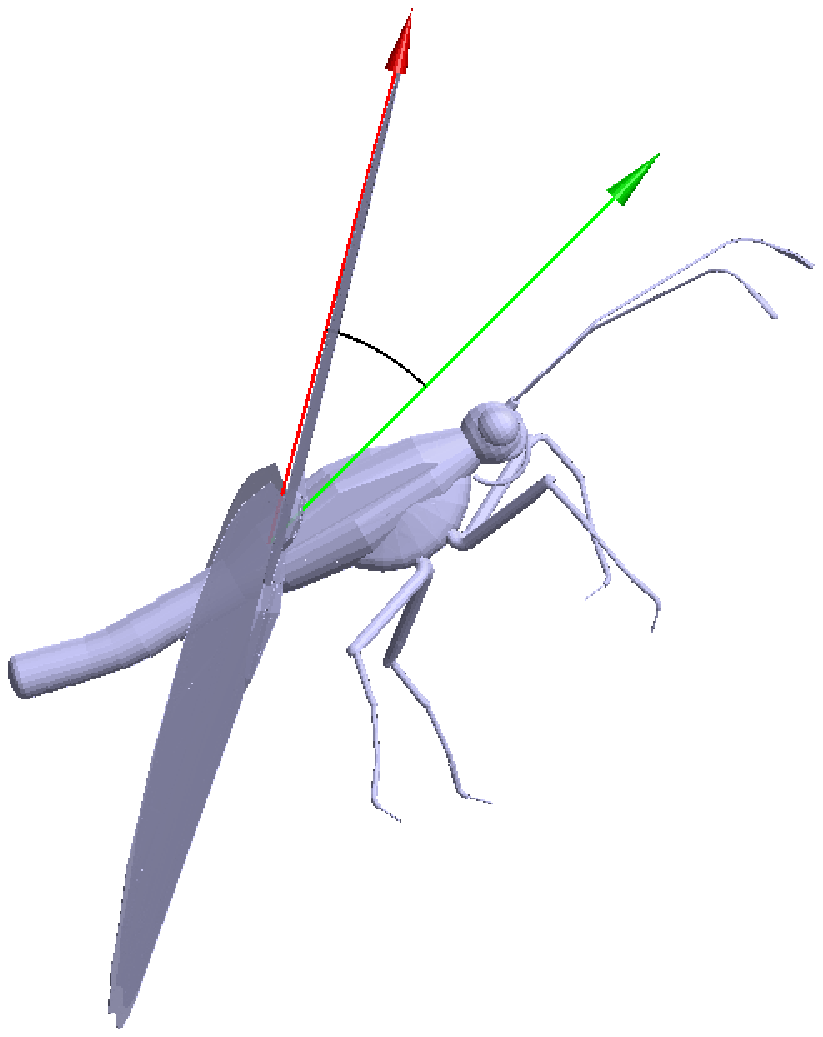}}
				\put(1.3,1.7){\scalefont{$\theta>0$}}
				\put(1.1,2.3){\scalefont{$\mathbf{r}_x$}}
				\put(2.0,1.85){\scalefont{$\mathbf{s}_x$}}
				\end{picture}
			}
			}
			\subfigure[deviation angle,\newline $\psi_R\in[-\pi,\pi)$]{
			\scalebox{0.5}{
				\begin{picture}(3,2.6)(0,0)
				\put(0,0){\includegraphics[trim={3cm 2cm 3cm 2cm},clip,width=0.3\textwidth]{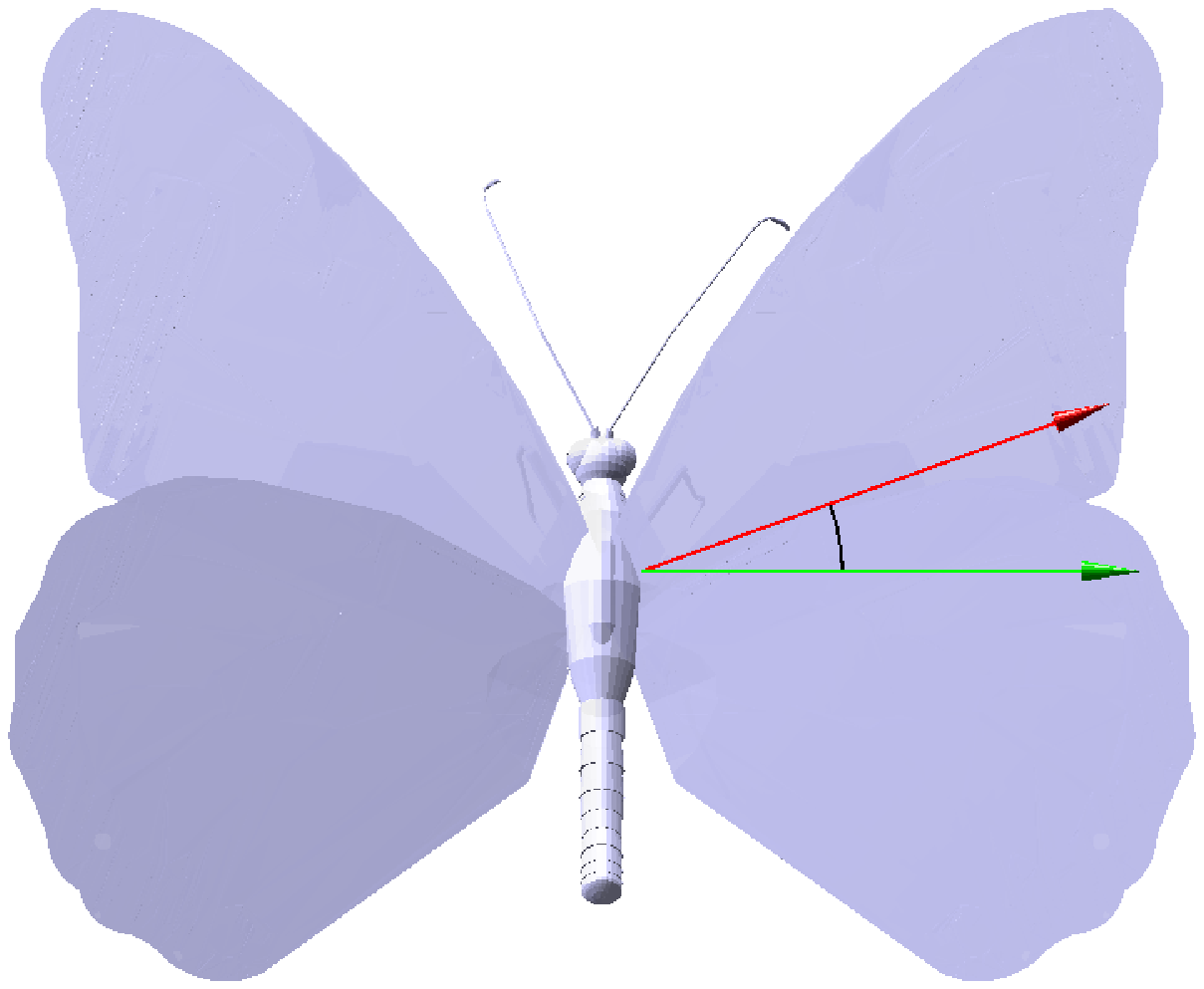}}
				\put(1.8,1.3){\scalefont{$\psi>0$}}
				\put(2.8,1.4){\scalefont{$\mathbf{r}_y$}}
				\put(2.8,0.95){\scalefont{$\mathbf{s}_y$}}
				\end{picture}
			}
			}
		}
		\caption{Euler angles \cite{sridhar2020geometric} : positive values are indicated from $ \mathcal{F}_S $ (green) to $ \mathcal{F}_R $ (red)}\label{fig:wing_Euler}
	\end{figure}
    The attitude of the right wing relative to $\mathcal{F}_S$, namely $Q_R\in\SO$ is described by 1--3--2 Euler angles $(\phi_R(t), \psi_R(t), \theta_R(t))$ (\Cref{fig:wing_Euler}) as
	\begin{align*}
	Q_R= \exp(\beta \hat e_2)\exp(\phi_R \hat e_1) \exp(-\psi_R \hat e_3) \exp(\theta_R\hat e_2),
	\end{align*}
	and its time-derivative is $ \dot Q_R = Q_R \hat \Omega_R $ for $\Omega_R\in\Re^3$. 

	\item \textit{Left Wing:}
        Similarly, for the left wing,
        \[ Q_L = \exp(\beta \hat e_2)\exp(-\phi_L \hat e_1) \exp(\psi_L \hat e_3) \exp(\theta_L \hat e_2) ,\]
        with the set of Euler-angles $(\phi_L(t), \psi_L(t), \theta_L(t))$, and
         $ \dot Q_L = Q_L \hat \Omega_L $ for $\Omega_L\in\Re^3$.

	\item \textit{Abdomen:}
	The abdomen is considered as a rigid body attached to the body via a spherical joint.
	The frame fixed to the abdomen is $\mathcal{F}_A=\{\mathbf{a}_x, \mathbf{a}_y, \mathbf{a}_z\}$,
    and its attitude relative to the body is denoted by $Q_A\in\SO$ with $\dot Q_A = Q_A\hat\Omega_A$ for $\Omega_A\in\Re^3$. 
	When there is no rotation relative to the body, its orientation is identical to $\mathcal{F}_B$. 
\end{itemize}

\subsection{Wing Kinematics}

The motion of the wing relative to the body is referred to as wing kinematics. 
As described above and depicted in \Cref{fig:wing_Euler}, it is defined by three angles, referred to as the flapping angle, the pitch angle, and the deviation angle.  
Here we adopt the particular wing kinematics model presented in~\cite{berman2007energy}.

Let $f\in\Re$ be the frequency of flapping in $\mathrm{Hz}$ and $T=\frac{1}{f}$ be the period in seconds. 
The flapping angle is given by a smoothed triangular waveform,
\begin{align}
\phi(t) & = \frac{\phi_m}{\sin^{-1} \phi_K}\sin^{-1}(\phi_K\cos(2\pi f t)) + \phi_0,\label{eqn:phi}
\end{align}
where $\phi_m\in\Re$ is the amplitude, $\phi_0\in\Re$ is the offset, and $0 < \phi_K \leq 1$ determines waveform shape (sinusoidal when $\phi_K\rightarrow 0$; triangular when $\phi_K\rightarrow 1$).
The initial pitch angle is $\phi(0)=\phi_m+\phi_0$.
During $0\leq t \leq \frac{T}{2}$, it decreases to $\phi(\frac{T}{2}) = -\phi_m+\phi_0$. 
According to the sign convention illustrated at \Cref{fig:wing_Euler}, it represents an upstroke when $0\leq t \leq \frac{T}{2}$, and a downstroke when $\frac{T}{2} \leq t \leq T$.

The pitch angle is given by a hyperbolic function,
\begin{align}
\theta(t) = \frac{\theta_m}{\tanh \theta_C} \tanh( \theta_C \sin(2\pi f t + \theta_a)) +\theta_0,\label{eqn:theta}
\end{align}
where $\theta_m\in\Re$ is the amplitude of pitching, $\theta_0\in\Re$ is the offset, $\theta_C\in(0,\infty)$ determines the waveform (sinusoidal when $\theta_C\rightarrow 0$; step function when $\theta_C\rightarrow \infty$), and $\theta_a\in(-\pi,\pi)$ describes phase offset (advance rotation for $\theta_a>0$; symmetric rotation for $\theta_a=0$; delayed rotation for $\theta_a<0$). The value of $\theta_C$ is related to the duration of wing pitch reversal. 

Finally, the deviation angle is given by 
\begin{align}
\psi(t) = \psi_m \cos(2\pi \psi_N f t + \psi_a) + \psi_0,\label{eqn:psi}
\end{align}
where $\psi_m\in\Re$ is the amplitude, $\psi_0\in\Re$ is the offset, and the parameter $\psi_a\in(-\pi,\pi)$ is the phase offset. 
The parameter $\psi_N$ is either 1 or 2: $\psi_N=1$ corresponds to one oscillation per a flapping period, and $\psi_N=2$ is for a figure-eight motion.

\subsection{Quasi-steady Aerodynamics}
Next, we present the expressions for the aerodynamic forces and moments generated by the wings.
Here, the blade-element theory~\cite{ellington1984aerodynamics} is utilized along with a quasi-steady aerodynamics assumption.
More explicitly, it implies that the aerodynamic force generated by an infinitesimal chord is independent of the span-wise velocity component,
and that the force and moment generated are equivalent to those for steady motion at the same instantaneous velocity and angle of attack.

Since the expression for the left wing can be similarly obtained, only the right wing is discussed below.
Consider an infinitesimal segment on the right wing at a distance $ r $ from the wing root.
In $ \mathcal{F}_R $, the relative velocity of the aerodynamic center of this chord element ($ c(r) $), projected to avoid span-wise components is
\begin{align}
    U_R(r) & = (I_{3\times 3}- e_2 e_2^T) Q_R^T( R^T (\dot x-v_{\mathrm{wind}}) + \Omega\times \mu_R )\nonumber \\
           & \quad + r  (Q_R\Omega + \Omega_R )\times e_2.\label{eqn:Ur}
\end{align}
where $v_{\mathrm{wind}}\in\Re^3$ is the uniform wind velocity in $\mathcal{F}_I$.
From this, the angle of attack $\alpha_R(r) \in [0,\frac{\pi}{2}]$ is
\begin{align}
\alpha_R (r) = 
\cos^{-1} ( \frac{|e_1^T U_R(r)|}{\|U_R(r)\|} ) = \sin^{-1} ( \frac{|e_3^T U_R(r)|}{\|U_R(r)\|} ) \label{eqn:alpha}
\end{align}
Let $\rho\in\Re$ be the atmospheric density, and let $C_L, C_D\in\Re$ as the lift and drag coefficients, respectively. 
The corresponding aerodynamic forces and moment generated by the infinitesimal wing segment are
\begin{align}
dL_R(r) &= \frac{1}{2}\rho U_R^2 C_L(\alpha) c\ \mathrm{sgn} (e_1^T U_R e_3^T U_R) \frac{e_2\times U_R}{\|e_2\times U_R\|} dr \nonumber \\
        &= \frac{1}{2}\rho  C_L(\alpha) c\ \mathrm{sgn} (e_1^T U_R e_3^T U_R) (e_2\times U_R)\|U_R\|  dr \label{eqn:dLR} \\
dD_R(r) &= - \frac{1}{2}\rho  C_D(\alpha(r)) c(r) \|U_R(r)\|U_R(r) dr, \label{eqn:dDR}\\
dM_R(r) & = r e_2 \times (dL_R + dD_R).
\end{align}
The total lift, drag, and moment of the right wing are obtained by integrating above span-wise for $r\in[0,l]$, for the right wing span $l\in\Re^3$.
Compared with the other models considering the uniform force over the wing, the above expression captures the span-wide variations of the angle of attack and the aerodynamic forces, which are critical for FWUAVs with relative large wings flapping at a low frequency. 

\subsection{Lagrangian Mechanics}

The configuration of the above flapping wing UAV model is described by $g=(x,R,Q_R,Q_L, Q_A)$, and as such, it is a mechanical system evolving on the Lie group $\G=\Re^3\times \SO^4$, whose Lie algebra is simply $\g = \Re^3 \times \so^4 \simeq \Re^3 \times (\Re^3)^4$.
Therefore, its equation of motion can be formulated by Lagrangian mechanics on a manifold~\cite{lee2017global}.

The kinematics equation on $\G$ is given by
\begin{equation}
\dot g = g \xi,\label{eqn:dot_g}
\end{equation}
for $\xi = (\dot x, \Omega, \Omega_R, \Omega_L, \Omega_A)\in\g$.
Let $\mathbf{J}:\G\times \g\rightarrow \g^*$ be a symmetric, positive-definite inertia tensor.
And define $(\mathbf{K}_g(\xi))(\cdot):\G\times \g\rightarrow \g^*$ such that
$ \T_e^* \L_g \cdot \D_g \mathbf{J}_g(\xi) \cdot \chi = (\mathbf{K}_g(\xi)) (\chi) = \mathbf{K}_g(\xi)\chi $. 
Consider a Lagrangian $L:\G\times \g \rightarrow \Re$ given by
$ L(g,\xi) = \frac{1}{2} \pair{ \mathbf{J}_g(\xi), \xi } - U(g) $
for a potential $U:\SO\rightarrow \Re$.
The corresponding Euler--Lagrange equations \cite{lee2017global} are written as
\begin{gather}
    \mathbf{J}_g(\dot \xi) + \mathbf{K}_g(\xi)\xi - \ad^*_\xi \cdot \mathbf{J}_g(\xi)  - \frac{1}{2}\mathbf{K}^*_g(\xi)\xi \nonumber \\
    + \T_e^* \L_g \D_g U(g) =\mathbf{f},\label{eqn:EL_full}
\end{gather}
where $\mathbf{f}\in\Re^{15}$ represents the effects of the aerodynamic forces and the internal torques at the joint. 
The explicit expressions for each term of the above equation are available in~\cite{sridhar2020geometric}. 

The above equation is driven by the control torque acting on each joint. 
Instead, it is assumed that the motion of the wings, abdomen, and body are prescribed, i.e., $(R(t),Q_R(t),Q_L(t),Q_A(t))$ are given as functions of time.
This is common in the literature, as the inertia of the wings are relatively small. 
Substituting these, the reduced equation of motion for the position is
\begin{align}
    m\ddot x & + 
\sum_{i\in\{R,L,A\}} \big\{ \mathbf{J}_{i_{12}} \dot\Omega + \mathbf{J}_{i_{13}}\dot\Omega_i 
+ \ \mathbf{K}_{i_{12}}\Omega + \mathbf{K}_{i_{13}}\Omega_i \big\}\nonumber \\
             & = R\sum_{i\in\{R,L,A\}} Q_i F_i + mg e_3, \label{eqn:mx_ddot}
\end{align}
where $F_i=L_i+D_i\in\Re^3$ denotes the resultant aerodynamic force at each component, $m\in\Re$ is the total mass, and $g\in\Re$ is the gravitational acceleration. 

In short, \eqref{eqn:mx_ddot} describes how the position of FWUAV evolves over time for a given wing kinematics and abdomen undulation, while capturing the dynamic coupling between components, and the aerodynamic forces and moments. 

\subsection{Monarch Butterfly Model}

For numerical analyses presented in the remainder of this paper, the wing morphological parameters and the mass properties are chosen to be similar with those of Monarch, and the specific values are summarized in~\cite{sridhar2020geometric}.
For the aerodynamic coefficients, we adopt the experimental results presented in~\cite{dickinson1999wing,sane2001control} as summarized below.
\begin{align*}
	C_L(\alpha) & = 0.225 + 1.58 \sin( (2.13\alpha^\circ - 7.2) \frac{\pi}{180}), \\
	C_D(\alpha) & = 1.92 - 1.55 \cos( (2.04 \alpha^\circ - 9.82 ) \frac{\pi}{180}),
\end{align*}
where $\alpha^\circ = \alpha\frac{180}{\pi}$. 
Next, the aerodynamic force generated by the body and the abdomen is ignored, i.e., $F_A=0$. 
This is reasonable as the projected surface area of the body and the abdomen is negligible compared with the wings. 
\section{Construction of Periodic Motion}\label{sec:periodic}

In this section, we construct a periodic motion for the above dynamic model. 
The objective is to find wing kinematics parameters such that a certain desired maneuver for the position dynamics is achieved. 
Here we focus on a hovering flight, i.e., at the end of one cycle of wing flapping the position and the velocity returns to the initial values so that its motion can be repeated. 

\subsection{Body/Abdomen Kinematics}

When generating a periodic motion, the body and the abdomen are assumed to undulate.
This is motivated by the flight characteristics of live Monarch butterfly exhibiting a nontrivial pitching motion of the body and the abdomen~\cite{sridhar2020geometric}.
Specifically, the attitude of the body is defined as $R(t) = \exp(\theta_B(t)\hat e_2)$, where the body pitch angle is 
\begin{align}
    \theta_B(t) = \theta_{B_m} \cos{(2 \pi f t + \theta_{B_a})} + \theta_{B_0},
\end{align}
for fixed parameters $\theta_{B_m}$, $\theta_{B_a}$, and $\theta_{B_0}\in\Re$.

Similarly, the attitude of the abdomen relative to the body is $Q_A(t) = \exp(\theta_A(t)\hat e_2)$, where
\begin{align}
    \theta_A(t) = \theta_{A_m} \cos{(2 \pi f t + \theta_{A_a})} + \theta_{A_0},
\end{align}
for $\theta_{A_m},\theta_{A_a},\theta_{A_0}\in\Re$.

\subsection{Problem Formulation}

An optimization problem to generate this motion is formulated as follows,
\begin{itemize}
    \item The objective function is the absolute amount of energy input into the system and corresponding power required:
        \begin{equation}\label{eqn:obj_func}
            J = w_1 \int_{0}^{T} |E(t)| dt + w_2 \int_{0}^{T} |\dot{E}(t)| dt,
        \end{equation}
        where $w_1,w_2\in\Re$ are positive weighting factors, and $E(t)\in\Re$ is composed of the kinetic energy and the gravitational potential energy:
        \begin{equation}
            E(t) = \frac{1}{2} m \norm{\dot x(t)}^2 - mge_3^T x(t).
        \end{equation}
    \item The optimization parameters are
        \begin{itemize}
            \item flapping frequency: $f$
            \item stroke plane angle: $\beta$
            \item wing kinematics: $(\phi_m, \phi_K, \phi_0)$, $(\theta_m, \theta_C, \theta_0, \theta_a)$, $(\psi_m, \psi_0, \psi_a)$
            \item body undulation: $(\theta_{B_m}, \theta_{B_0}, \theta_{B_a})$
            \item abdomen undulation  $(\theta_{A_m}, \theta_{A_0}, \theta_{A_a})$
            \item initial velocity: $\dot x(0)$
        \end{itemize}
    \item There are equality constraints to ensure periodicity: 
        \begin{equation}
            x(0) = x(T),\quad  \dot{x}(0) = \dot{x}(T).
        \end{equation}
        There is an inequality constraint for feasibly flapping:
        \begin{equation}
            |\phi_m| + |\phi_0| < \pi/2.
        \end{equation}
        All of the optimization parameters are bounded by prescribed limits.
\end{itemize}

\begin{figure}[h!]
	\centerline{
		\subfigure[Position $x$]{
			\includegraphics[width=0.5\linewidth]{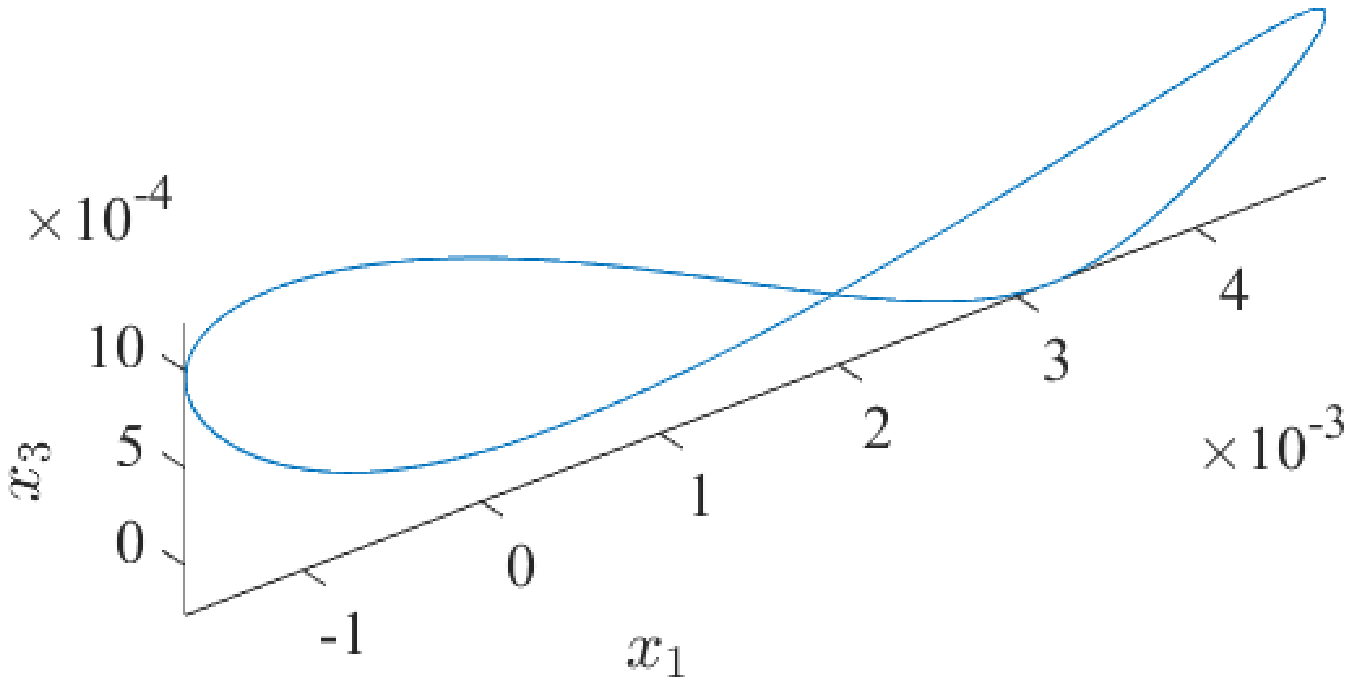}
		}
		\hfill
		\subfigure[Velocity $\dot x$]{
			\includegraphics[width=0.5\linewidth]{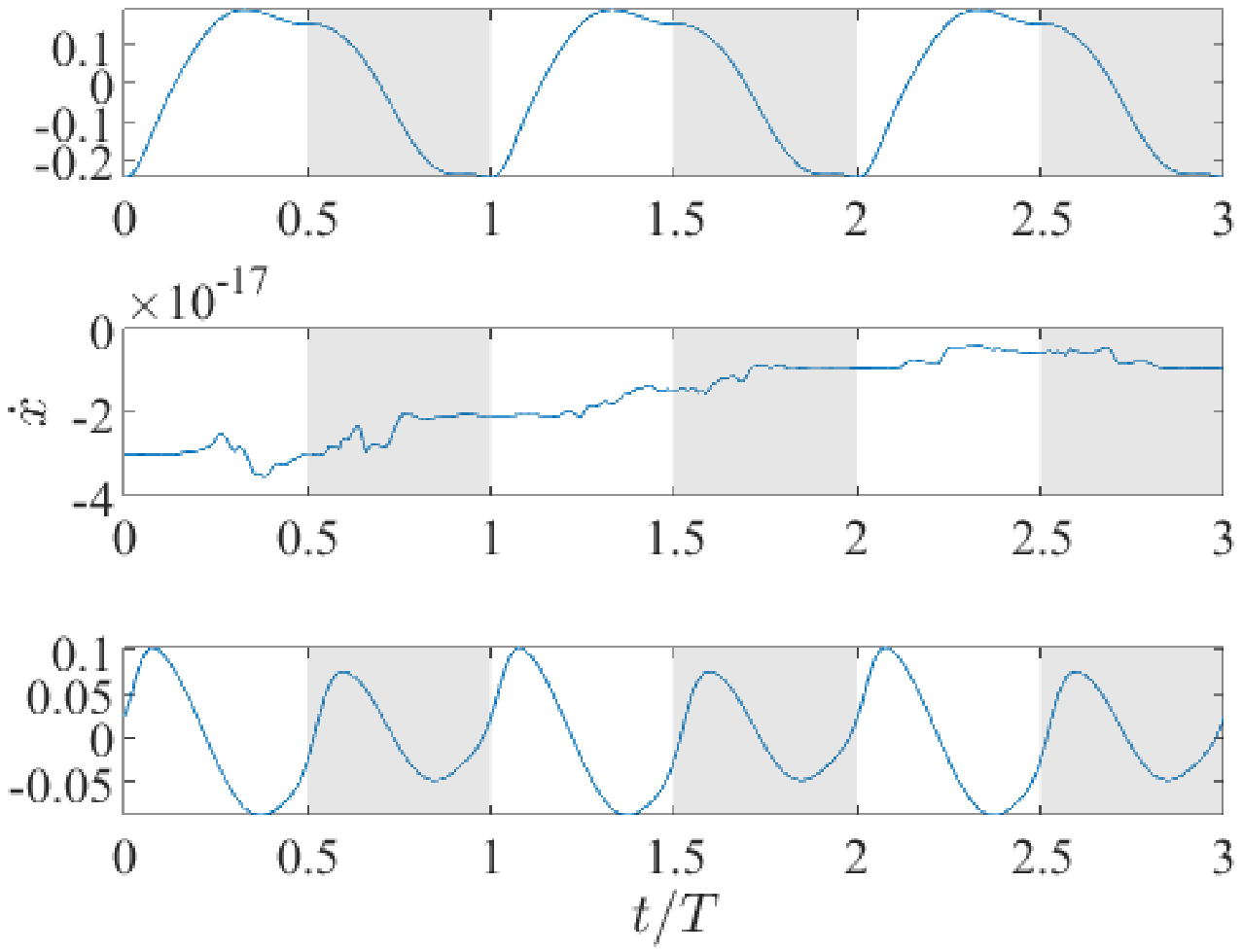}
		}
	}
	\centerline{
        \subfigure[Wing kinematics angles $(\phi,\theta,\psi)$\newline (in degrees)]{
			\includegraphics[width=0.5\linewidth]{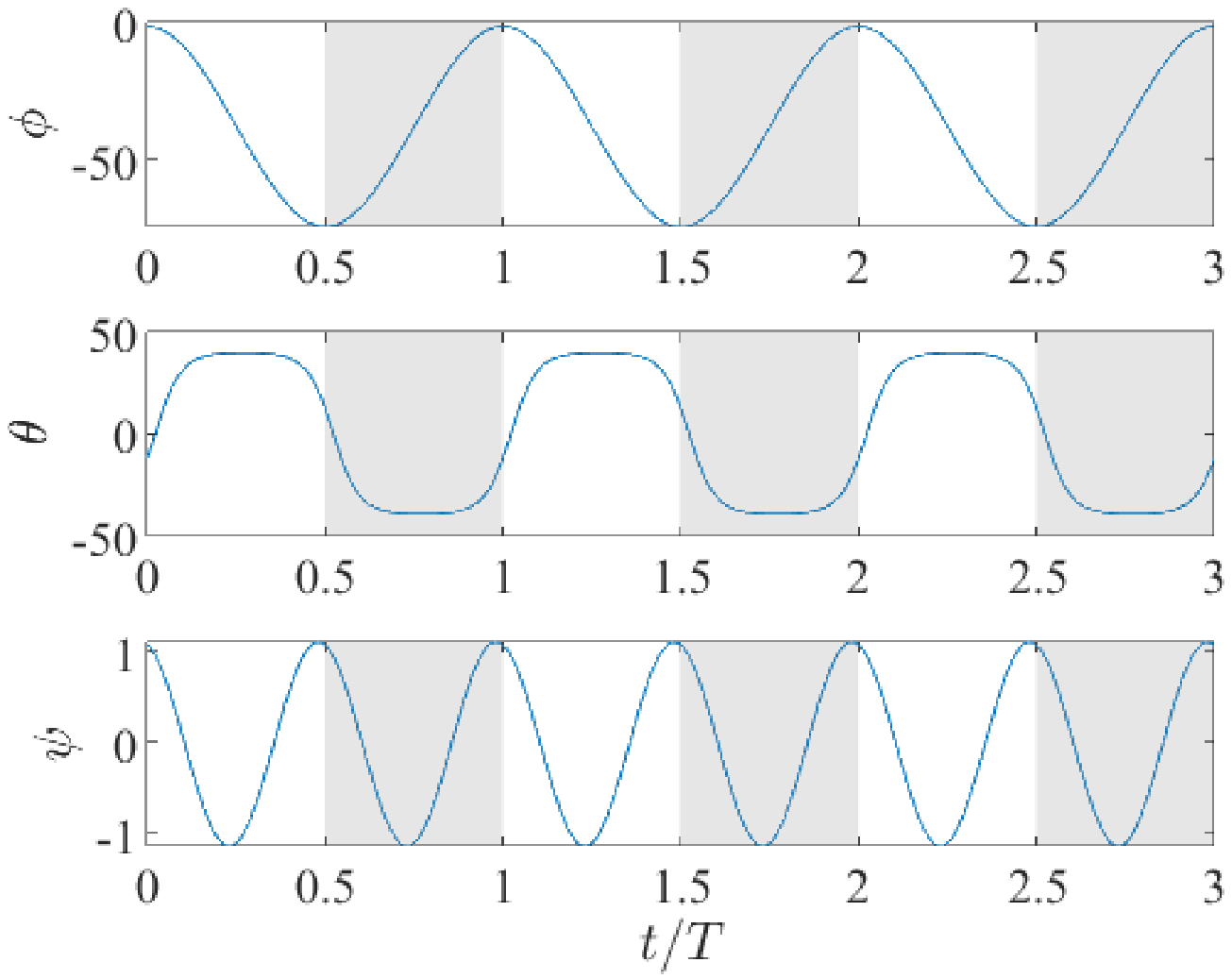}
		}
		\hfill
		\subfigure[Relative abdomen/ body pitch $\theta_B,\theta_A$ (in degrees)]{
			\includegraphics[width=0.5\linewidth]{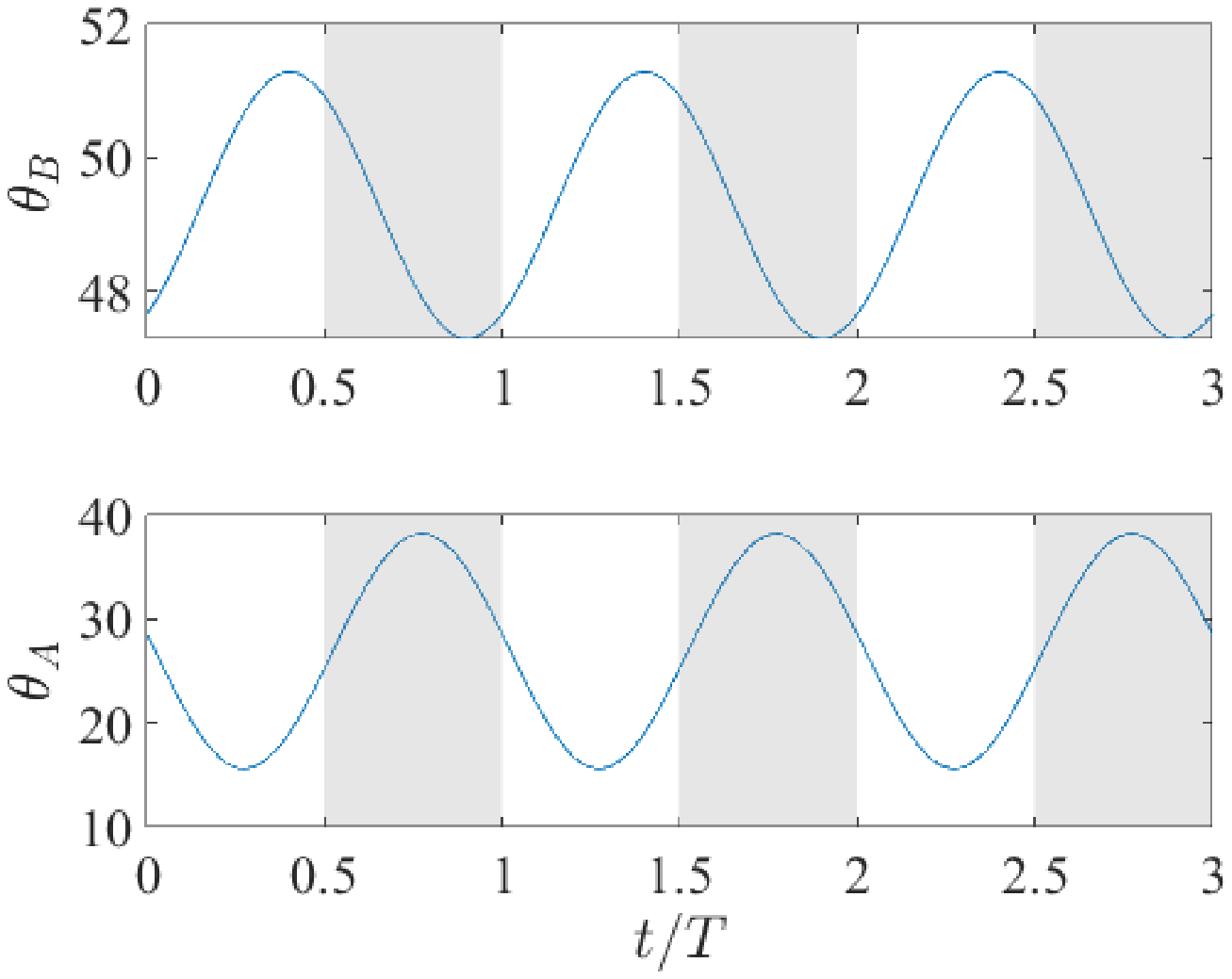}
		}
	}
    \caption{Hovering periodic orbit generated using optimized parameters; shaded region corresponds to downstrokes}
	\label{fig:hover_pos_vel}
\end{figure}

It is considered that the motion of the left wing is always symmetric to the right wing.
The above parameter optimization problem is solved via MATLAB global optimization tools, such as \texttt{multistart} or \texttt{particleswarm}.
The optimized parameters are summarized at \Cref{tab:hover_params}, and the corresponding periodic maneuver is illustrated at \Cref{fig:hover_pos_vel}.
\newcolumntype{m}{>{$}l<{$}} 
\begin{table}[h!]
	\caption{Optimized parameters}\label{tab:hover_params}
	\begin{center}
		\begin{tabular}{|m|m|m|}
			\hline
			\text{Parameters} & \text{With abdomen} & \text{Without abdomen}\\
			& \text{undulation} & \text{undulation} \\\hline
			f & 11.6689 & 11.7539 \\
			\beta & 0.7782 & 0.9396 \\
			\phi_m & 0.6355 & 0.6449 \\
			\phi_K & 0.2866 & 0.2618 \\
			\phi_0 & -0.6599 & -0.6635 \\
			\theta_m & 0.6893 & 0.6977 \\
			\theta_C & 2.1703 & 2.3602 \\
			\theta_0 & 0.0098 & 0.0034 \\
			\theta_a & -0.1410 & -0.1737 \\
			\psi_m & 0.0196 & 0.0057 \\
			\psi_N & 2 & 2 \\
			\psi_0 & -0.0003 & 0.0033 \\
			\psi_a & 0.2506 & 0.3517 \\
			\theta_{B_m} & 0.0348 & 0.0014 \\
			\theta_{B_0} & 0.8602 & 0.7202 \\
			\theta_{B_a} & -2.5204 & -0.3049 \\
			\theta_{A_m} & 0.1970 & \text{\textemdash\textemdash} \\
			\theta_{A_0} & 0.4696 & -0.1712 \\
			\theta_{A_a} & 1.4270 & \text{\textemdash\textemdash} \\
			\dot x_1(0) & -0.2458 & -0.2816 \\
			\dot x_2(0) & 0.0000 & 0.0000 \\
			\dot x_3(0) & 0.0230 & 0.0200 \\\hline
			\text{Optimized}\ J & 0.0409 & 0.0504 \\\hline
		\end{tabular}\\[0.1cm]
		($ f_{natural} = 10.2247\,\si{Hz} $ and $ \psi_N $ is fixed at $ 2 $)
	\end{center}
\end{table}

\subsection{Effects of Abdomen in Energy and Power}

\begin{figure}
	\centerline{
		\subfigure[Energy]{
			\includegraphics[width=0.5\linewidth]{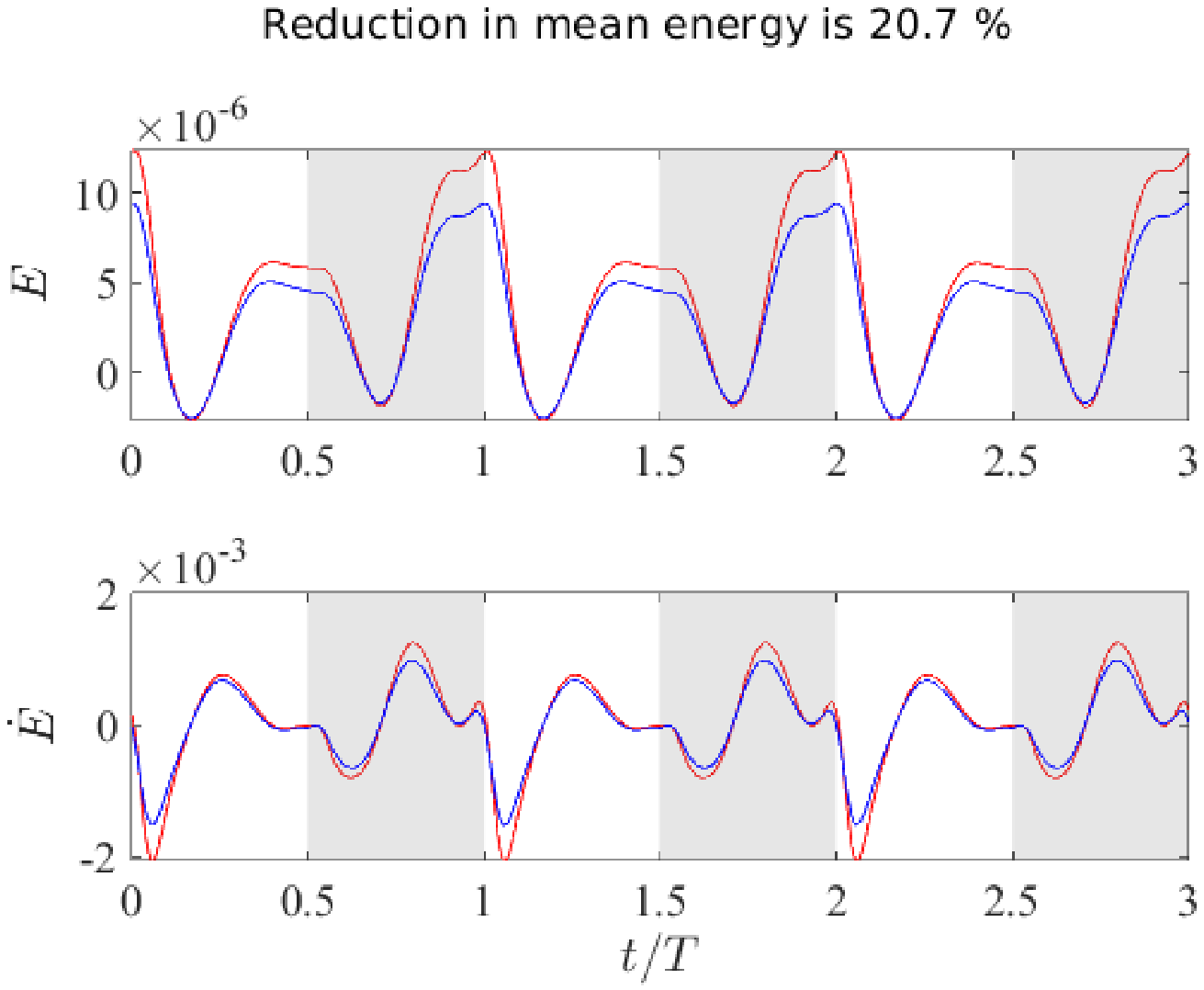}
		}
		\hfill
		\subfigure[Power]{
			\includegraphics[width=0.5\linewidth]{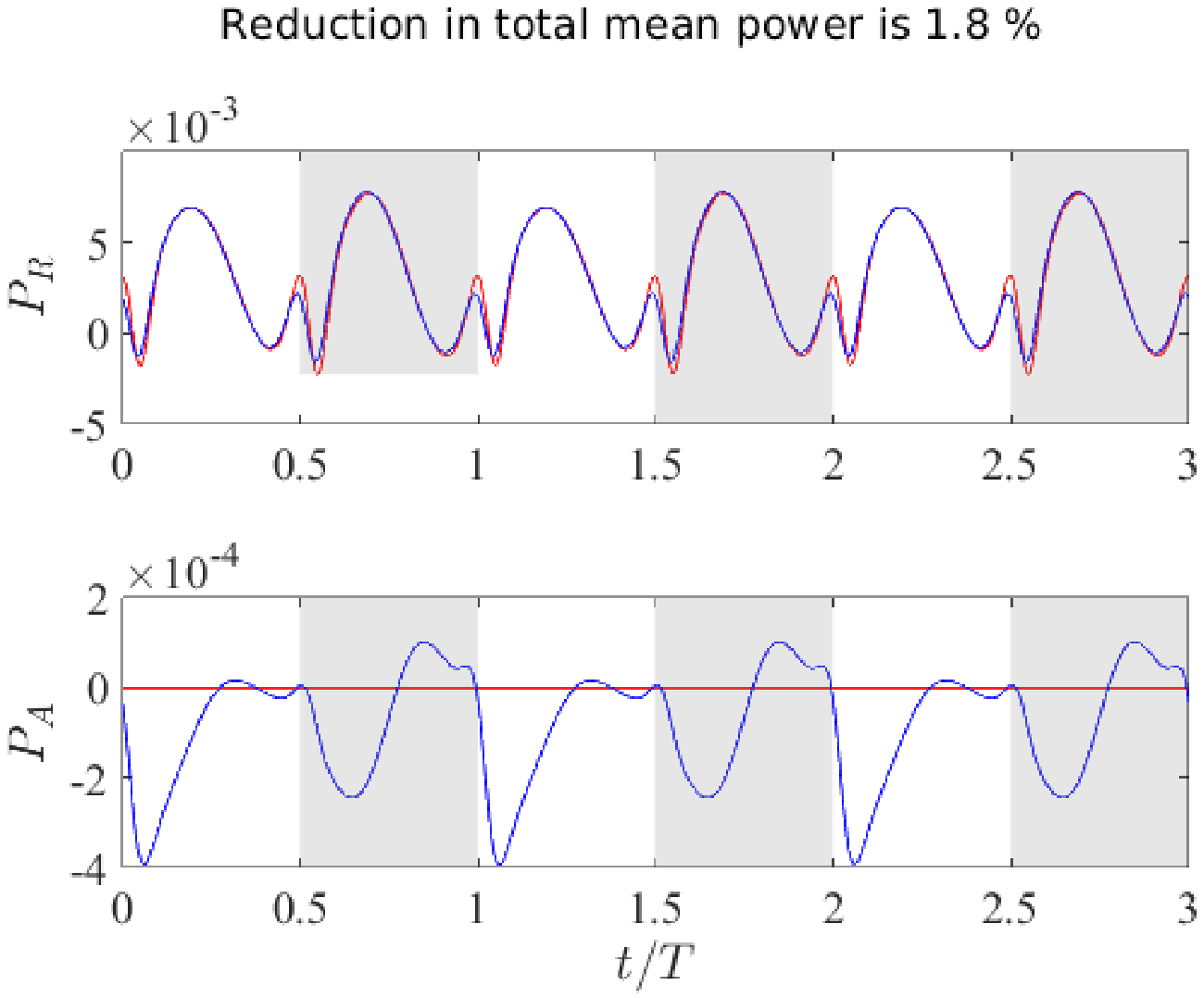}
		}
	}
	\centerline{
		\subfigure[Torque]{
			\includegraphics[width=0.5\linewidth]{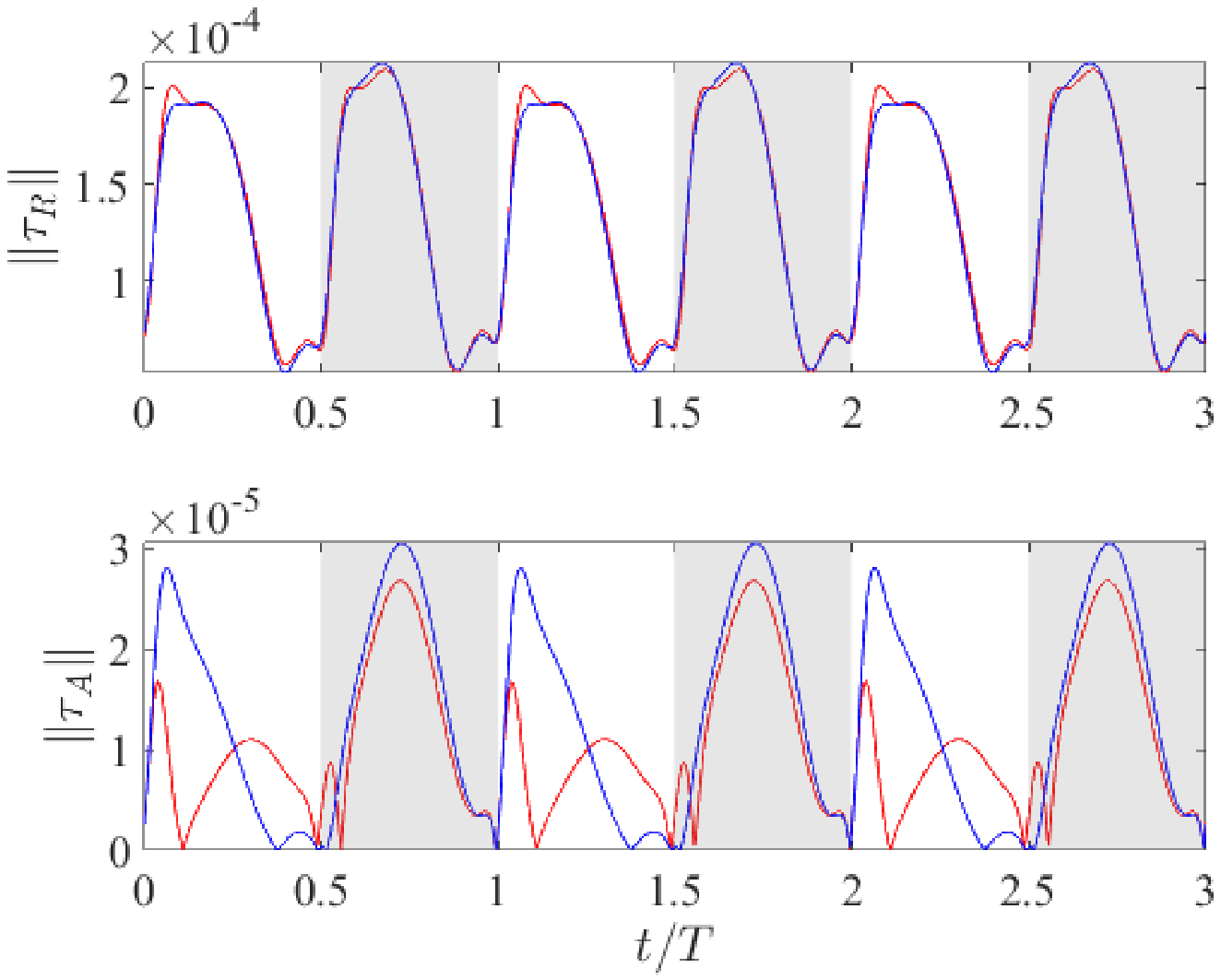}
		}
		\subfigure[Modeled torque with abdomen flapping]{
			\includegraphics[width=0.5\linewidth]{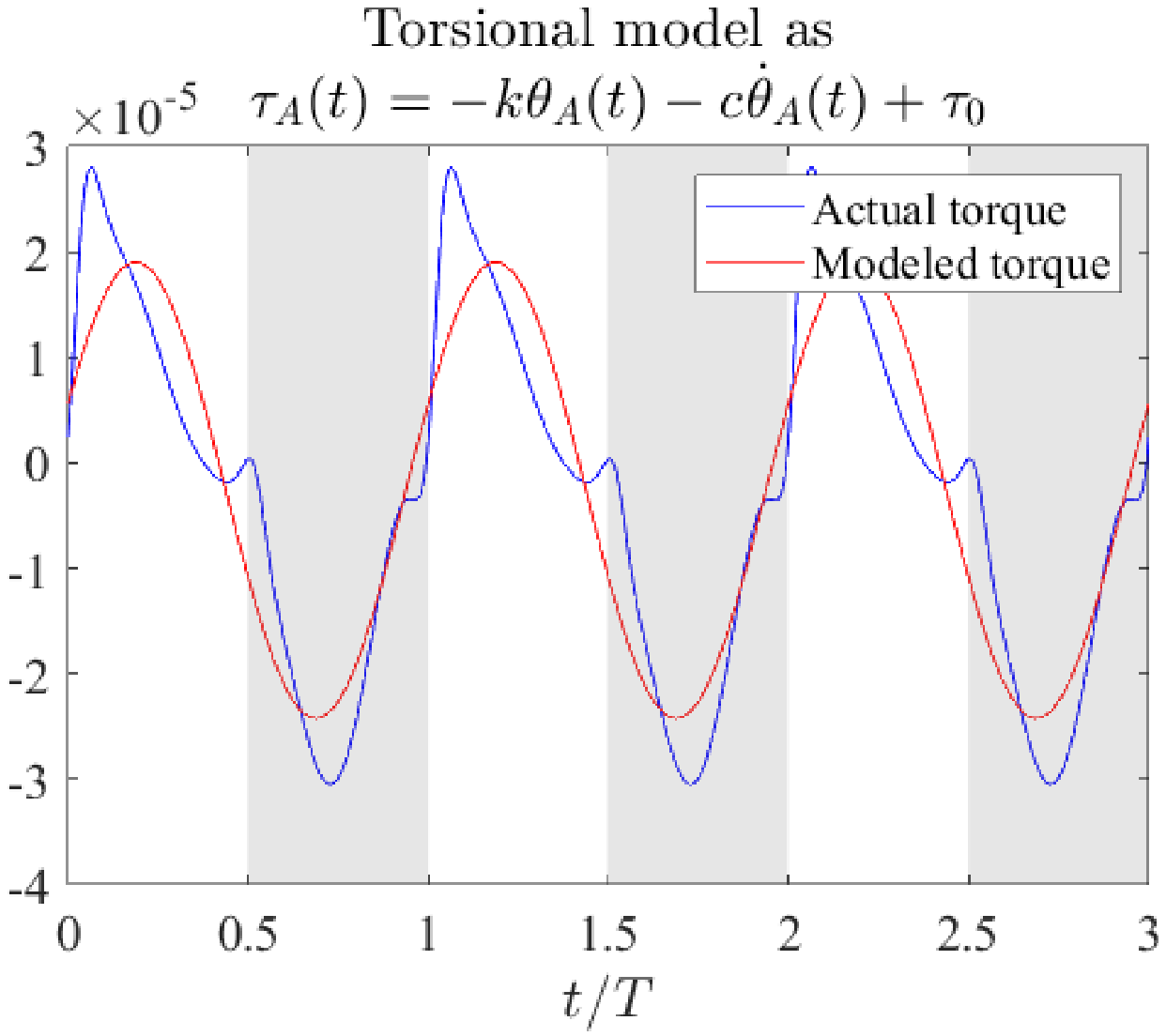}
		}
	}
    \caption{Comparison between hovering with abdomen undulation (blue) and hovering without  abdomen undulation (red); energy and power input to the model is relatively low if there is abdomen movement, i.e., $1.8 \%$ reduction in total mean power and $ 20.7 \%$ reduction in mean energy}
	\label{fig:hover_comp_ab}
\end{figure}

To further study the effects of abdomen, another optimization is carried out while assuming that the abdomen does not undulate relative to the body, i.e., $ \theta_A(t) = \theta_{A_0} $ for a fixed parameter $\theta_{A_0}$.
The resulting optimized parameters are summarized at the second column of \Cref{tab:hover_params}, where the optimal value of the objective function is increased by $23\%$ when the abdomen is fixed. 
The corresponding variation of $E$ and $\dot E$ are presented at \Cref{fig:hover_comp_ab}.(a).
It is shown that the abdomen undulation reduces the variation of the total energy and its time-derivatives. 

Next, the torque required at the joint of the wing $(\tau_R,\tau_L)$ and the torque at the joint connecting the body and the abdomen $\tau_A$ are reconstructed from \eqref{eqn:EL_full}.
The corresponding power at the joints are computed as $P_R = \tau_R^T (Q_R \Omega_R)$, $P_A = \tau_A^T (Q_A \Omega_A)$.
The change of power over a flapping period is illustrated at \Cref{fig:hover_comp_ab}.(b), which shows $1.8\%$ reduction of the power with abdomen undulation. 

The torque at abdomen for the case with relative abdomen undulation can be modeled by a torsional spring/ damper at the joint. Considering $ \tau_A(t) = -k\theta_A(t) - c\dot{\theta}_A(t) + \tau_0 $ and using a least squares fit for the data, we obtain $ k = \num{9.4262e-05},\ c = \num{7.6634e-07},\ \tau_0 = \num{4.1641e-05}$ (see the subfigure (d)).
These show that the abdomen undulation may reduce the total energy and the power consumption required for hovering.
Interestingly, such beneficial effects may be achieved by allowing the abdomen to passively undulate after the muscle connecting the body and the abdomen is modeled as a torsional spring/damper. 

\section{Feedback Control of Flapping-Wing UAV}\label{sec:control}

In this section, we study the open-loop stability of the hovering flight acquired in the preceding section, and then, we propose a feedback control to improve the stability properties of the corresponding periodic orbit. 
The flapping frequency constructed from the morphological parameters of Monarch is $f=\SI{10.22}{\hertz}$, which is relatively lower than the flapping frequencies of other FWUAVs ranging up to $\SI{200}{\hertz}$. 
As such, the common approaches relying on the linearized dynamics averaged over a flapping cycle are ill-suited.
Here we study the stability of the periodic orbit using Floquet stability theory, and design a feedback control system accordingly.~\footnote{The Matlab software utilized for dynamics and control of the flapping wing model is available at \href{https://github.com/fdcl-gwu/FWUAV}{\texttt{https://github.com/fdcl-gwu/FWUAV}}.}

\subsection{Floquet Stability}

Let $\mathbf{x}\in\Re^6$ be the congregated state of the position and the velocity, i.e., $\mathbf{x}=(x, \dot x) \in\Re^6$. 
Also, let $\mathbf{x}_d = (x_d, \dot x_d)\in\Re^6$ be the periodic orbit for hovering constructed in the preceding subsection, satisfying $\mathbf{x}_d(t+T) = \mathbf{x}_d(t)$ for any $t>0$.

The perturbation of the state $\mathbf{x}$ from the periodic orbit $\mathbf{x}_d$ is denoted by $\delta\mathbf{x}= \mathbf{x}-\mathbf{x}_d=(\delta x, \delta \dot x)\in\Re^6$. 
Ignoring the higher-order terms, the evolution of the perturbation is described by the equation of motion, namely \eqref{eqn:mx_ddot} linearized about $\mathbf{x}_d$ as
\begin{align}
    \frac{d (\delta x)}{dt} &= \delta \dot{x} \label{eqn:del_x_dot}\\
m \frac{d (\delta \dot{x} )}{dt} &= R(t)  \{ Q_R(t) \delta F_R(t,\delta \dot x(t)) + Q_L(t) \delta F_L(t,\delta \dot x(t))\}, \label{eqn:pos_vel_pert}
\end{align}
where $R(t),Q_R(t),Q_L(t)$ corresponds to the body undulation and the wing kinematics for the periodic orbit, and $\delta F_R, \delta F_L$ represent the variation of the aerodynamic force due to the perturbation of the velocity. 
For example, 
\begin{align}
    \delta F_R &= \delta L_R + \delta D_R, \label{eqn:delta_F_R}
\end{align}
where $\delta L_R, \delta D_R$  can be computed from \eqref{eqn:dLR}, \eqref{eqn:dDR} and
\begin{align*}
\delta C_L(\alpha(r)) &= 1.58 \cos( (2.13\alpha^\circ - 7.2) \frac{\pi}{180}) \times 2.13\ \delta\alpha_R(r), \\
\delta C_D(\alpha(r)) &= 1.55 \sin( (2.04 \alpha^\circ - 9.82 ) \frac{\pi}{180}) \times 2.04\ \delta\alpha_R(r),\\
\delta U_R(r) &= (I_{3\times 3}- e_2 e_2^T) Q_R^T R^T \delta\dot x, \\
\delta \alpha_R(r) &= \frac{-1}{\sin(\alpha_R(r))} \mathrm{sgn}(e_1^T U_R(r)) \\
                   & \quad \times e_1^T\left(I_{3\times 3} -  \frac{U_R(r)U_R^T(r)}{\norm{U_R(r)}^2}\right) \frac{\delta U_R(r)}{\norm{U_R(r)}}.
\end{align*}

Utilizing these, the linearized equations of motion, \eqref{eqn:del_x_dot}, \eqref{eqn:pos_vel_pert} can be rearranged into the following matrix form: 
\begin{equation}\label{eqn:periodic_ode}
    \delta \dot{\mathbf{x}} = \mathbf{A}(t) \delta\mathbf{x},
\end{equation}
where the matrix $ \mathbf{A}(t)\in\Re^{6\times 6} $ is periodic with the period $T$.
While the solution of \eqref{eqn:periodic_ode}, namely $\delta\mathbf{x}(t)$ is not periodic in general, it can be written as a linear combination of a set of periodic solutions multiplied by so-called \textit{characteristic multiplier}~\cite{teschl2012ordinary}.

More specifically, let $\Psi(t)\in\Re^{6\times 6}$ be the solution of the matrix differential equation $\dot\Psi = \mathbf{A}\Psi$, starting from a nonsigular $\Psi(0)\in\Re^{6\times 6}$.
The matrix $\Psi(t)$ is referred to as the \textit{fundamental matrix} of \eqref{eqn:periodic_ode}. 
Using the fact that $\mathbf{A}(t)$ is periodic, one can show that $\Psi(t+T)$ is also a fundamental matrix, and there exists a non-singular constant matrix $\mathbf{M}\in\Re^{6\times 6}$, referred to as the \textit{monodromy matrix}, such that $\Psi(t+T) = \Psi(t)\mathbf{M}$ for all $t$. 
One of the simplest way to compute the monodromy matrix is $\mathbf{M} = \Psi(T)$ with $\Psi(0)=I_{6\times 6}$.
In the subsequent development, it is assumed that $\Phi(0)=I_{6\times 6}$ for simplicity. 

Let the $i$-th pair of the eigenvalue and the eigenvector of the monodromy matrix$\mathbf{M}$ be $(\rho_i, \mathbf{v_i})\in\Re\times\Re^6$, and suppose they are real. 
The eigenvalues $\rho_i$ of the monodromy matrix is referred to as the \textit{characteristic multiplier} of \eqref{eqn:periodic_ode}.
Define $\delta\mathbf{x}_i(t)\in\Re^6$ be the solution of \eqref{eqn:periodic_ode} starting from $\delta\mathbf{x}(0)=\mathbf{v}_i$, i.e., $\delta\mathbf{x}_i (t) = \Psi(t) \mathbf{v}_i$. 
Then, we have
\begin{align}
    \delta\mathbf{x}_i (t+T) & = \Psi(t+T) \mathbf{v}_i = \Psi(t) \mathbf{M} \mathbf{v}_i = \rho_i \Psi(t) \mathbf{v}_i \nonumber\\
                             & = \rho_i \delta\mathbf{x}_i(t).
\end{align}
In other words, the solution of \eqref{eqn:periodic_ode} starting from the eigenvector of the monodromy matrix is scaled by the corresponding eigenvalue after each period. 
As the general solution of $\delta\mathbf{x}(t)$ starting from an arbitrary initial condition can be written as a linear combination of $\{\delta\mathbf{x}(t)\}_{i=1}^6$,
the periodic orbit is attractive if $|\rho_i|<1$ for all $i$.

This further provides the shape of the characteristic mode. 
Let $\mu_i\in\Re$ be defined such that $\rho_i = e^{\mu_i T}$, and $\{\mu_i\}_{i=1}^6$ is referred to as the \textit{characteristic exponent} of \eqref{eqn:periodic_ode}.
Define $\mathbf{p}_i(t) = \delta\mathbf{x}_i (t)e^{-\mu_i t }$.
It is straightforward to show $\mathbf{p}_i(t)$ is periodic, as $\mathbf{p}_i(t+T) = \rho_i \delta \mathbf{x}_i(t) e^{-\mu_i T} e^{-\mu_i t}=\mathbf{p}_i(t)$.
Therefore, 
\begin{align*}
    \delta\mathbf{x}_i (t) = e^{\mu_i t} \mathbf{p}_i(t),
\end{align*}
which states that the solution of \eqref{eqn:periodic_ode} is composed of a set of periodic orbits exponentially scaled by the exponential of the characteristic exponent multiplied by the time. 
These are readily generalized for complex eigenvalues and vectors, using the fact that they appear in a complex conjugate pair. 

\subsection{Stability of Hovering Flight}

The characteristic multiplier of the hovering flight constructed in Section \ref{sec:periodic} are as follows. 
\begin{align}\label{eqn:char_mul}
    \rho & = \lbrace 1,\, 1,\, 1,\, 0.3763,\, 0.6234,\, 0.5023 \rbrace,
\end{align}
and the corresponding eigenvectors are 
\begin{align*}
    &[\mathbf{v}_1,\ldots \mathbf{v}_6]\\
    & = 
    \begin{bmatrix}
        1     & 0    &     0  &  0.0081  &  -0.0278  &  0\\
        0  &  1   &      0  &  0 &  0 &   0.1326\\
        0    &     0  &  1 &   0.0003 &  -0.1888  &  0\\
        0    &     0     &    0  & -0.9998  &  0.6337  & 0\\
        0    &     0     &    0   &      0  &       0  & -0.9912\\
        0    &     0     &    0  & 0.0183  &  0.7496 &  0
    \end{bmatrix}.
\end{align*}


Since  $ |\rho_i| \le 1 $, the periodic orbit of hovering is stable.
The first three characteristic mode with $\rho=1$ corresponds to the perturbation of the position. 
Since the aerodynamic forces are not altered by $\delta x$ in~\eqref{eqn:delta_F_R}, it is not surprising that the characteristic modes for position displacements have the characteristic multiplier of one.

On the other hands, the perturbation along the velocity diminishes asymptotically. 
As the last three components of $(\mathbf{v}_4,\mathbf{v}_5,\mathbf{v}_6)$ span $\Re^3$, any velocity perturbation can be written as a linear combination of those three asymptotically stable modes with $|\rho_4|,|\rho_5|,|\rho_6| < 1$. 
The fourth mode and the fifth mode have the velocity perturbation within the longitudinal plane, along the forward direction $\delta\dot x_1$ and along the vertical direction $\delta\dot x_3$,
where the fourth mode is dominated by the velocity perturbation along the forward direction. 
The last mode is along the lateral direction. 
These results are consistent with~\cite{sridhar2020geometric}, where it is reported that the velocity trajectory of the forward climbing flight of Monarch is asymptotically stable. 
The asymptotic stability in velocity can be deduced from the fact that the aerodynamic forces are proportional to the angle of attack and the velocity squared.
For example, if the FWUAV is pulled downward, it increases the effective angle of attack and the velocity at the wing, thereby generating an additional lift to push it back.

\subsection{Control System Design}

Next, we design a feedback control system to asymptotically stabilize the periodic orbit of hovering.
The objective is to stabilize the position modes and to improve the stability properties of the velocity modes. 
The control input corresponds to the wing kinematics parameters and the abdomen undulation, which are adjusted to the current position and the velocity that are assumed to be available. 

\begin{figure}[h!]
	\centerline{
		\subfigure[Longitudinal forces]{
			\includegraphics[width=\linewidth]{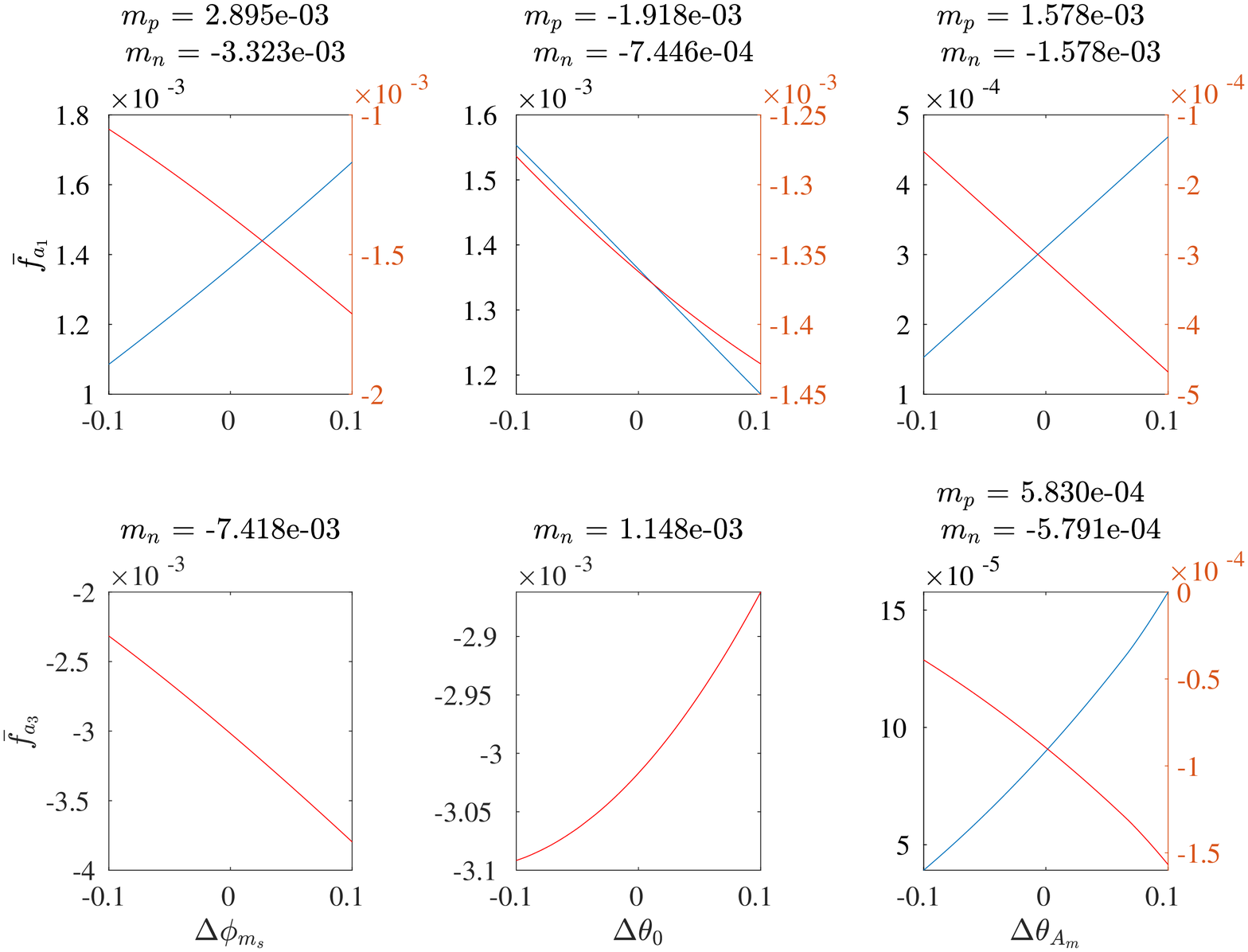}
		}
	}
	\centerline{
		\subfigure[Lateral force]{
			\includegraphics[width=0.4\linewidth]{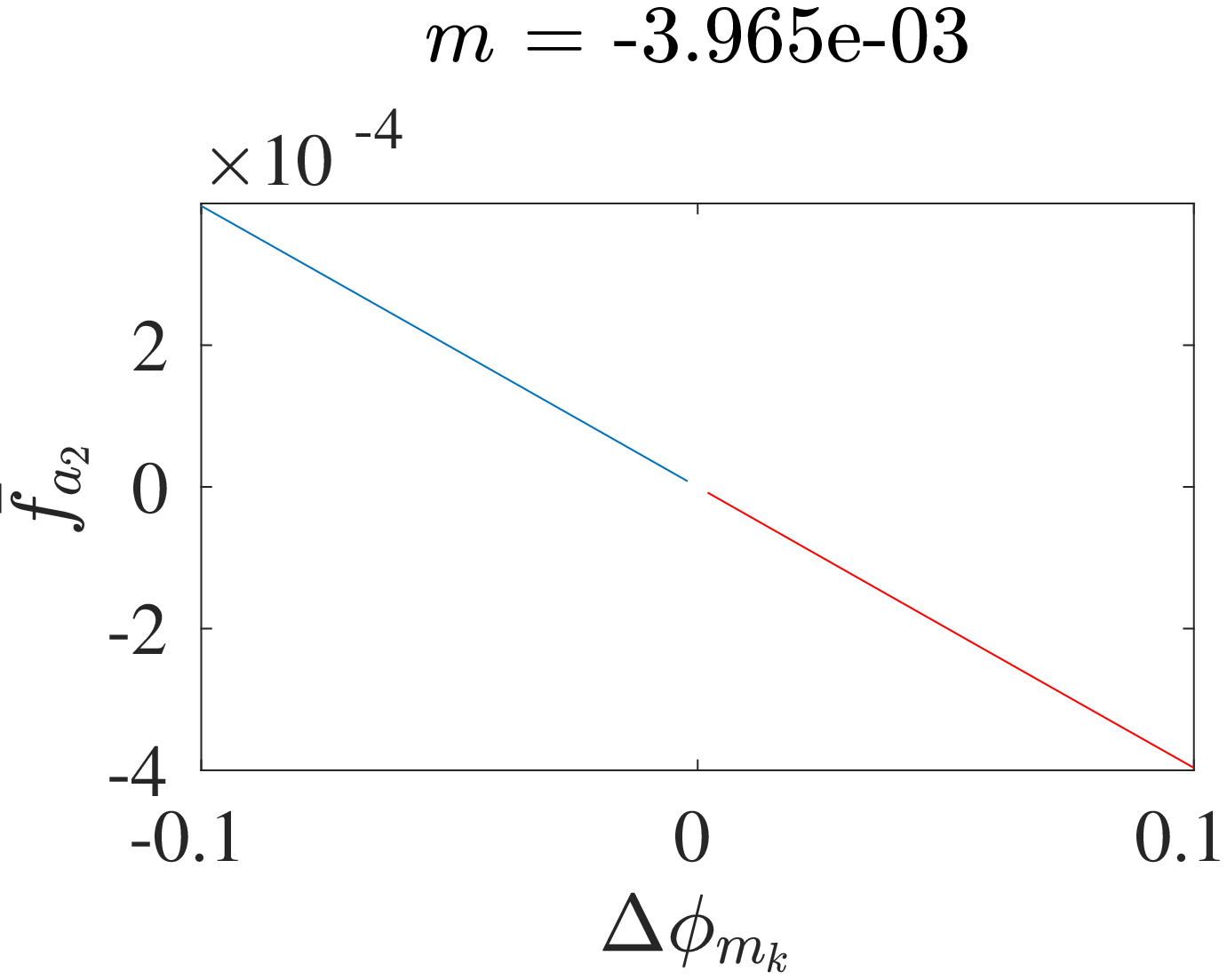}
		}
	}
	\caption{Effects of control parameters on the aerodynamic forces: mean of positive values (blue) and mean of negative values (red)}
	\label{fig:hover_param}
\end{figure}

First, we study the relation between these wing kinematic angles, as modeled in~\eqref{eqn:phi}, \eqref{eqn:theta}, \eqref{eqn:psi}, and the aerodynamic forces acting on the body.
Let the resultant aerodynamic force and the coupling effects of the abdomen undulation be $f_a\in\Re^3$,
\begin{align}
    f_a = R (Q_RF_R+ Q_LF_L) -(\mathbf{J}_{A_{13}} \dot\Omega_A + \mathbf{K}_{A_{13}} \Omega_A).\label{eqn:f_a}
\end{align}
We choose the following four control parameters:
\begin{itemize}
    \item $\Delta\phi_{m_s} = (\Delta\phi_{m,R} + \Delta\phi_{m,L}) /2$: the mean of the flapping amplitude of the right wing and the left wing for overall magnitude of the aerodynamic force
    \item $\Delta\phi_{m_k} = (\Delta\phi_{m,R} - \Delta\phi_{m,L}) /2$: the difference of the flapping amplitude to generate a lateral force
    \item $\Delta\theta_0$: the shifts in the pitching angle of both wings to rotate the direction of the aerodynamic force in the longitudinal plane
    \item $\Delta\theta_{A_m}$: the amplitude of the abdomen undulation
\end{itemize}
\Cref{fig:hover_param} illustrates the variation of $f_a$ averaged over a flapping frequency, namely $\bar f_a\in\Re^3$ with respect to the variations of the above parameters. 
We take the average of the positive values of $f_a$, separately from the negative values. 
For example, the mean of the positive values of the first element of $f_a$ is constructed as follows. 
Let $ T_p = \braces{t \in [0, T]\ \vert\ f_{a_{1}}(t) > 0} $.
We have
\begin{align*}
    \bar f_{a_{1,p}} = \tfrac{\int_{t\in T_p} f_{a_1}(\tau) d\tau }{ \int_{t\in T_p} d\tau}.
\end{align*}
The mean of the negative values $\bar f_{a_1,n}$ are defined similarly. 
The reason is that the variation of the control parameters affects $f_{a_1}$ in the opposite way, depending on the sign of $f_{a_1}$. 
These are summarized at~\Cref{fig:hover_param}, with the corresponding slope for the positive values, namely $m_p$, and the slope $m_n$ for the negative values. 

Employing these relations, we choose the wing kinematics parameters such that the variation of the averaged force $\Delta\bar f_a$ matches with a PID controller as follows.
\begin{equation} \label{eqn:PID_controller}
    \Delta \bar f_a = m (K_P \delta x (t) + K_D \delta\dot x(t) + K_I \int_{0}^{t} \delta x(\tau) d\tau),
\end{equation}
where $ K_P, K_D, K_I>0 $ are the controller gains.
After decomposing the longitudinal motion from the lateral motion, the wing kinematics parameters are computed from the following linear approximation.
For the longitudinal forces, we have
\begin{align}
    S
    \begin{bmatrix}
    \Delta \phi_{m_s} \\
    \Delta \theta_0 \\
    \Delta \theta_{A_m}
    \end{bmatrix} =&
    \begin{bmatrix}
    \Delta \bar f_{a_1} \\
    \Delta \bar f_{a_3}
    \end{bmatrix} ,\label{eqn:delta_phi_m}
\end{align}
where the matrix $S\in\Re^{3\times 3}$ is defined as
\begin{align}
    S =&
    \begin{bmatrix}
        \dfrac{\partial \bar f_{a_1}}{\partial \phi_{m_s}} & \dfrac{\partial \bar f_{a_1}}{\partial \theta_0} & \dfrac{\partial \bar f_{a_1}}{\partial \theta_{A_m}} \\[1em]
        \dfrac{\partial \bar f_{a_3}}{\partial \phi_{m_s}} & \dfrac{\partial \bar f_{a_3}}{\partial \theta_0} & \dfrac{\partial \bar f_{a_3}}{\partial \theta_{A_m}}
    \end{bmatrix}.\label{eqn:S}
\end{align}
When calculating $S$, the value of the sensitivity is chosen according to the sign of $\bar f_{a_i}$ as discussed above. 
For example, when $f_{a_1}>0$, we use $ \dfrac{\partial \bar f_{a_1}}{\partial \phi_{m_s}} = \dfrac{\partial \bar f_{a_{1,p}}}{\partial \phi_{m_s}} $.
As \eqref{eqn:delta_phi_m} is underdetermined, the control parameters are obtained by the minimum norm solution as
\begin{equation*}
\begin{bmatrix}
\Delta \phi_{m_s} \\
\Delta \theta_0 \\
\Delta \theta_{A_m}
\end{bmatrix} =
S^T (S S^T)^{-1}
\begin{bmatrix}
\Delta \bar f_{a_1} \\
\Delta \bar f_{a_3}
\end{bmatrix}.
\end{equation*}

For a comparison, we consider another case when the abdomen is not actively controlled, i.e., $\Delta\theta_{A_{m}}=0$.
The corresponding active control parameters $\Delta\phi_{m_s},\Delta\theta_0$ are computed by inverting the square matrix composed of the first two columns of \eqref{eqn:S}. 
These ensure that the resultant aerodynamic forces in the longitudinal plane match with \eqref{eqn:PID_controller}.

Next, for the lateral force,  
\begin{align}
	\Delta \phi_{m_k} &= \left( \frac{\partial \bar f_{a_2}}{\partial \phi_{m_k}} \right)^{-1} \Delta \bar f_{a_2}. \label{eqn:inverse_design}
\end{align}

\subsection{Stability of Controlled Hovering Flight}

\begin{figure}[h!]
	\centerline{
		\subfigure[Position]{
			\includegraphics[width=0.5\linewidth]{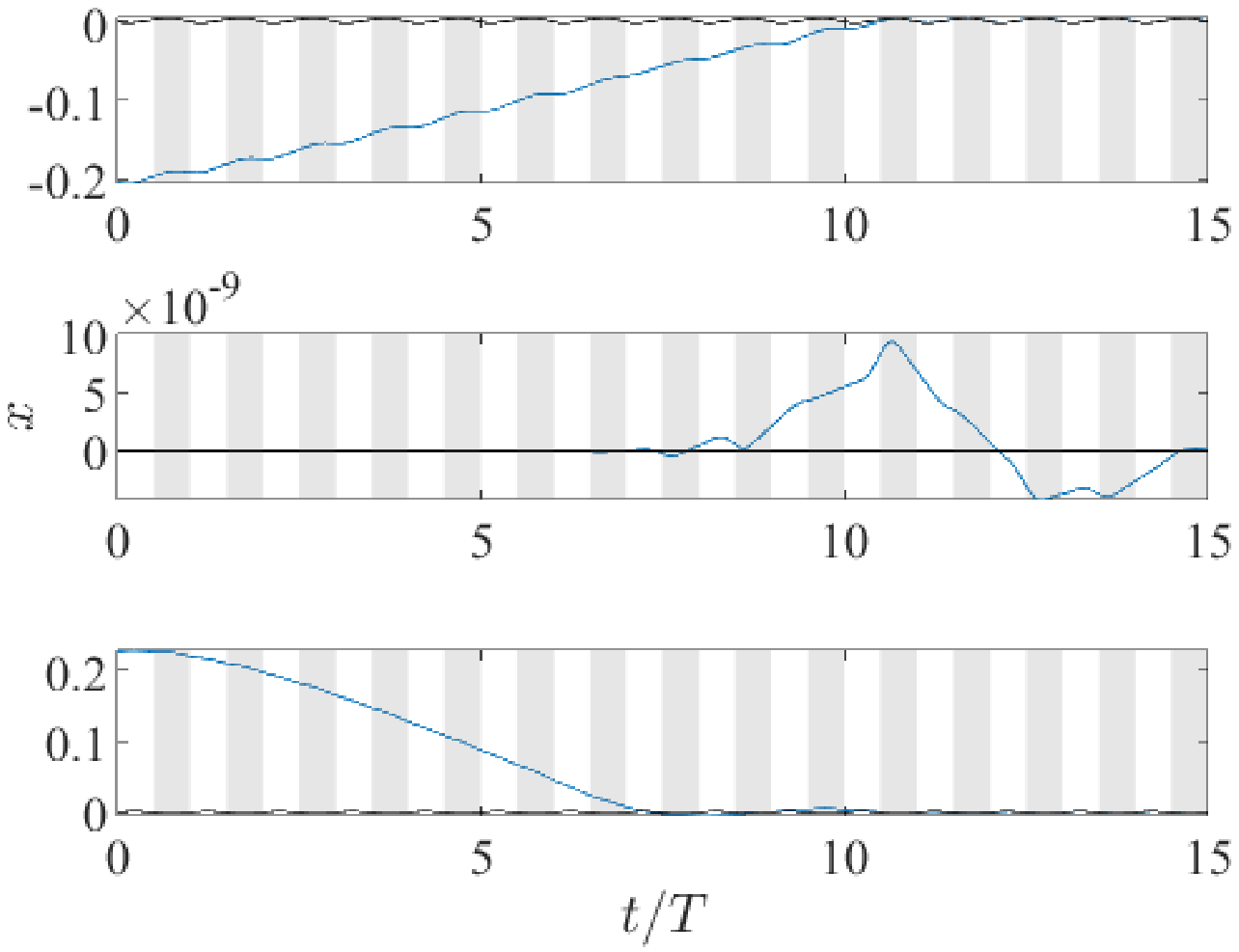}
		}
		\hfill
		\subfigure[Velocity]{
			\includegraphics[width=0.5\linewidth]{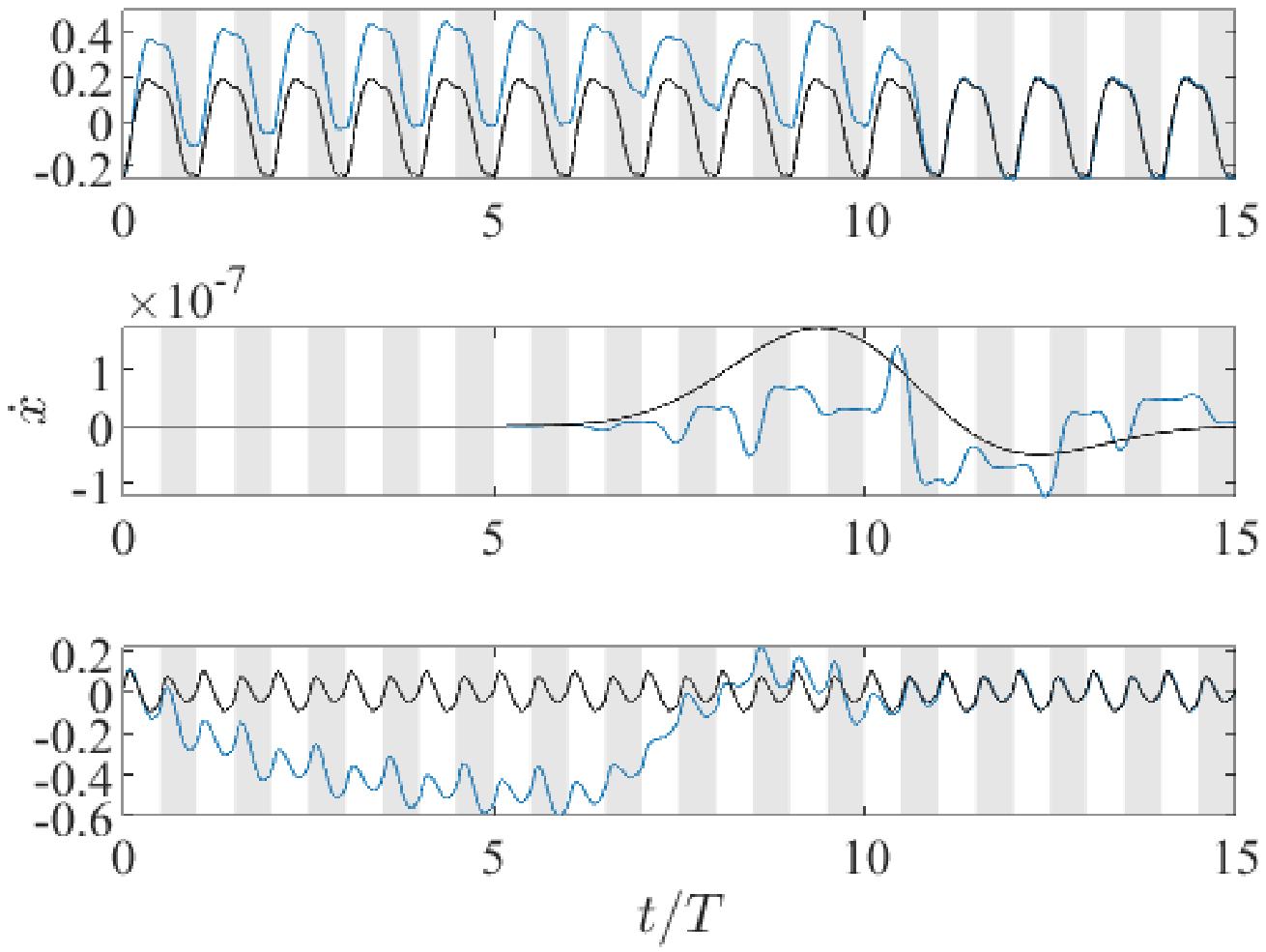}
		}
	}
	\centerline{
		\subfigure[3D position trajectory]{
			\includegraphics[width=0.5\linewidth]{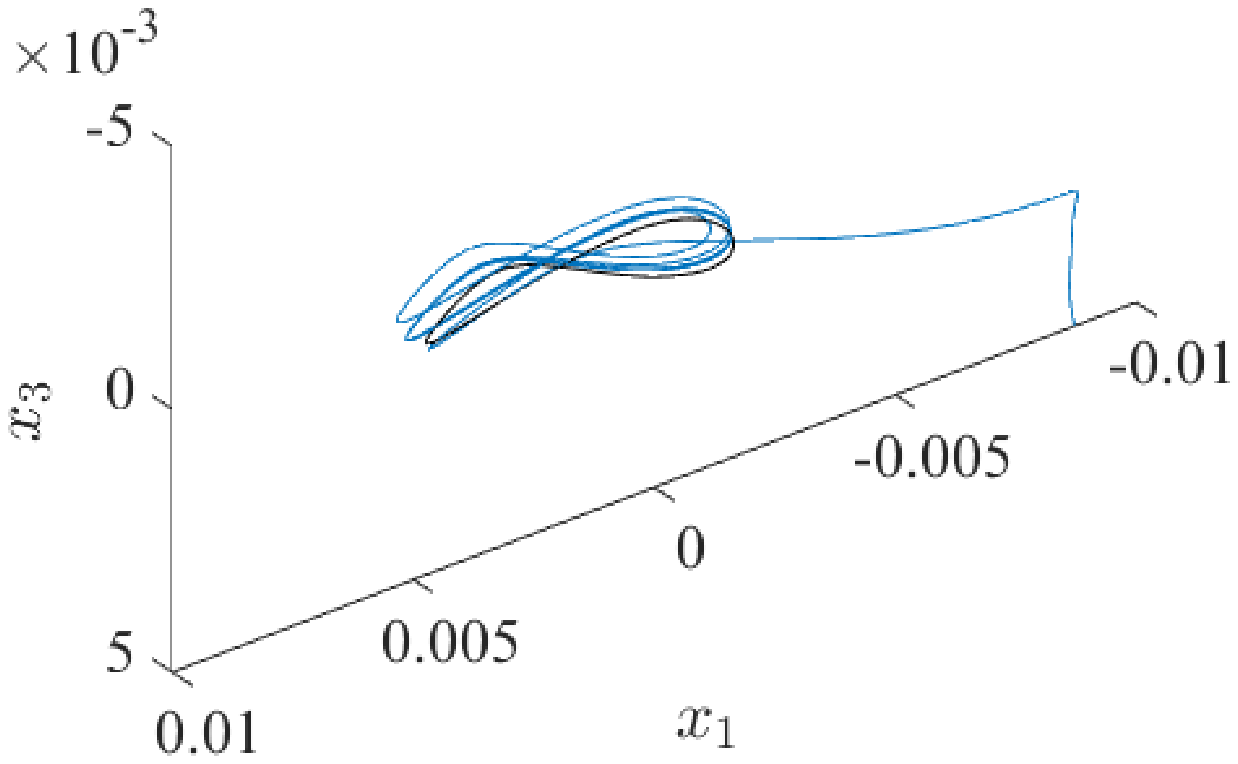}
		}
		\subfigure[Control inputs]{
			\includegraphics[width=0.5\linewidth]{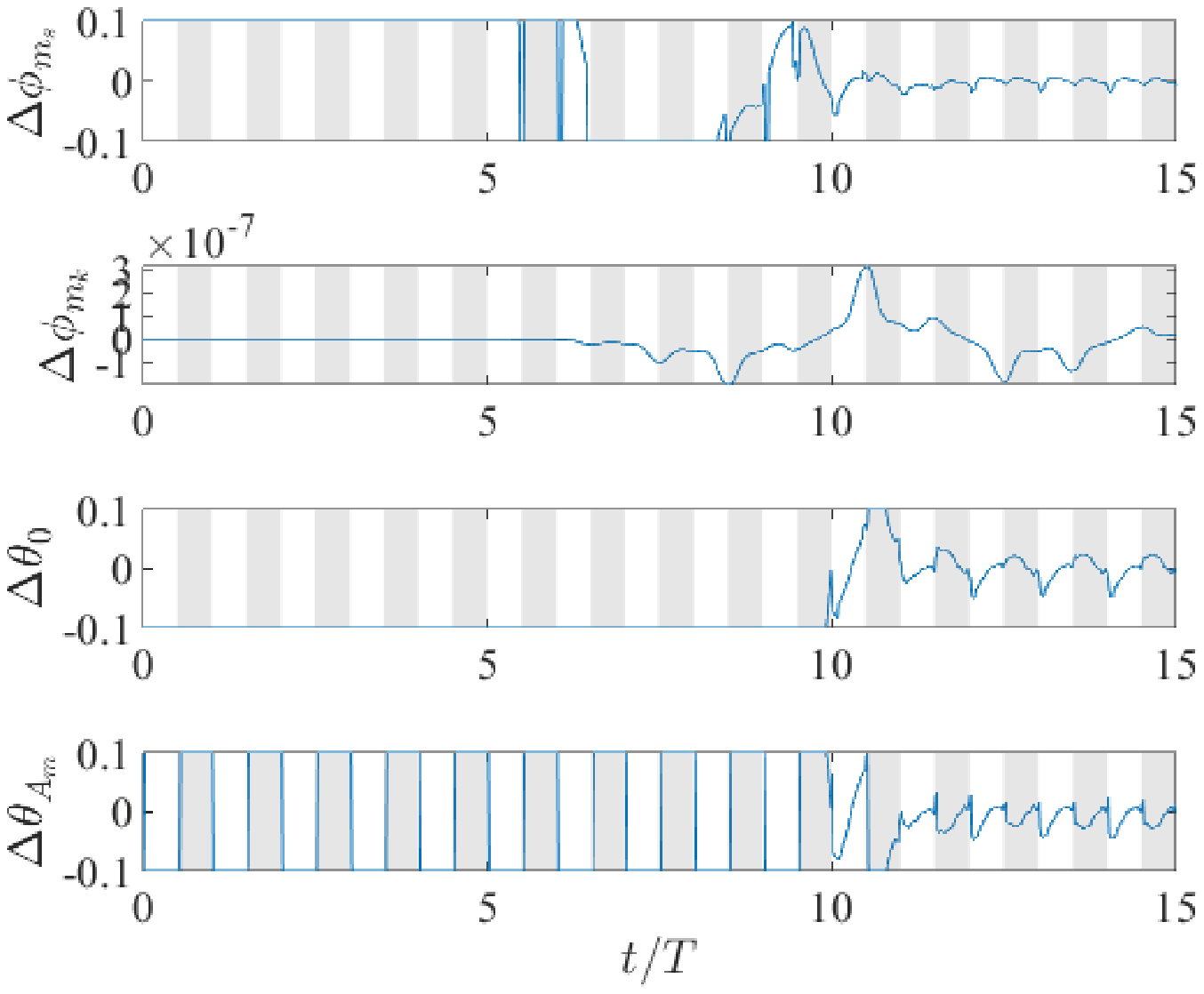}
		}
	}
	\caption{Position, velocity and control input trajectory obtained through the controlled dynamics: actual (blue), ideal (black); corresponding animated simulation can be found at \href{https://youtu.be/tMmmaVAm5D0}{\texttt{https://youtu.be/tMmmaVAm5D0}}}
	\label{fig:hover_control}
\end{figure}

In this subsection, we present simulation results with active abdomen control. 
The controller gains are chosen as $ K_P = 421.88, K_D = 15.60, K_I = 1.26 $.
Corresponding roots of the characteristic equation $\lambda^3 + K_D \lambda^2 + K_P \lambda + K_I$ are $-7.8 + 19i, -7.8 - 19i, -0.003 $.
\Cref{fig:hover_control} illustrates the position and the velocity trajectory over the controlled dynamics starting from a perturbed initial condition, and it is shown that the states asymptotically converge to the desired periodic orbit. 

Next, we verify the stability of the periodic orbit of the controlled dynamics via Floquet theory. 
Define an additional state $\delta I_x \in\Re^3$ for the integral term as $ \delta I_x = \int_0^t \delta x(\tau) d\tau$.
And, the controlled dynamics of \eqref{eqn:mx_ddot}, where $Q_i,\Omega_i$ are dependent of $\delta x$ and $\delta \dot x$, are numerically linearized.
This yields the system matrix $\mathbf{A}(t)\in\Re^{9\times 9}$ in \eqref{eqn:periodic_ode}, from which the monodromy matrix is computed.
The corresponding characteristic multipliers are complex numbers with the following magnitudes: 
\begin{align*}
    |\rho| = \lbrace & 1,\ 1,\ 1,\ 0.1186,\ 0.1186,\\
    & 0.2835,\ 0.2835,\ 0.6326,\ 0.6326 \rbrace.
\end{align*}
Since the magnitude of all of the characteristic multipliers is less than or equal 1, the periodic orbit for the proposed controlled dynamics is  stable.
Moreover, the characteristic modes for the first three with $\rho=1$ predominantly consist of the integral term.
Next, the modes with $ |\rho| = \braces{0.1186, 0.2835} $ correspond to the longitudinal dynamics and they are much smaller than those of uncontrolled ones, namely $ |\rho| = \braces{0.3763, 0.6234} $ in~\eqref{eqn:char_mul}, thereby illustrating an improvement in stability.
Finally, $ |\rho| = \braces{0.6326} $ corresponding to the lateral mode is greater than $ \rho = \braces{0.5023} $ of~\eqref{eqn:char_mul}.
However, it guarantees asymptotic stability of both of the position and the velocity dynamics. 

\subsection{Effects of Abdomen in Control}
\begin{figure}[h!]
	\centering
	\includegraphics[width=0.8\linewidth]{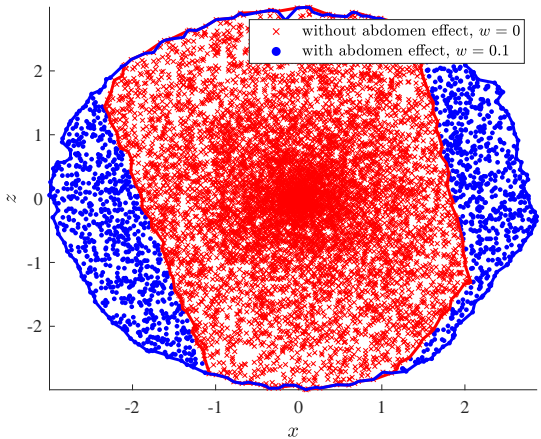}
	\caption{Initial position errors which yield converging trajectories through control design; corresponding bounded curves depict region of convergence}
	\label{fig:hover_mc_roa}
\end{figure}

Finally, we study the effects of the abdomen in the stability of the controlled dynamics. 
We consider two cases depending on the abdomen is actively controlled or not, and  for each case we estimate the region of attraction with respect to the initial position error in the longitudinal plane. 
More specifically, we select $ 10,000 $ random initial position errors of the form $ \delta x(0) = [e_x, 0, e_z] $, where $(e_x, e_z)$ are sampled from the uniform distribution on the circle represented by   $ \lbrace (r\cos{\theta}, r\sin{\theta})\ |\ (r, \theta) \in (0, 3) \times (0, 2\pi) \rbrace $.
Each random initial point is propagated through the controlled dynamics, and it is determined to be converged if the position and the velocity error from the desired periodic orbit becomes less than $ \num{1e-4} $ within 100 periods. 
\Cref{fig:hover_mc_roa} represents the set of initial conditions from which the controlled trajectory converged, and it is illustrated that the active abdomen undulation increases the region of attraction substantially. 

\begin{figure}[h!]
	\centering
	\includegraphics[width=0.7\linewidth]{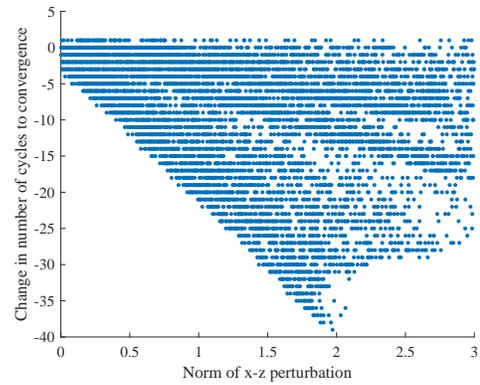}
	\caption{Comparison between control with and without abdomen}
	\label{fig:hover_mc_perf}
\end{figure}

Next, for the initial conditions that yield convergence for both cases, we compare the rate of convergence.
It is observed that the number of flapping cycles required for convergence is smaller when the abdomen is actively controlled. 
The corresponding reduction of the flapping cycles cased by active abdomen control is illustrated at~\Cref{fig:hover_mc_perf}, with respect to the varying magnitude of the initial error. 
For a wide variety of perturbations, it is found that there is a significant reduction.

In summary, the abdomen undulation can be actively controlled to improve the controller performance both by enlarging the region of attraction and by improving the rate of convergence.

\bibliographystyle{ieeetr}
\bibliography{root.bib}

\begin{thebibliography}{10}

\bibitem{sun2007dynamic}
M.~Sun, J.~Wang, and Y.~Xiong, ``Dynamic flight stability of hovering
  insects,'' {\em Acta Mechanica Sinica}, vol.~23, no.~3, pp.~231--246, 2007.

\bibitem{taha2012flight}
H.~E. Taha, M.~R. Hajj, and A.~H. Nayfeh, ``Flight dynamics and control of
  flapping-wing mavs: a review,'' {\em Nonlinear Dynamics}, vol.~70, no.~2,
  pp.~907--939, 2012.

\bibitem{sun2014insect}
M.~Sun, ``Insect flight dynamics: stability and control,'' {\em Reviews of
  Modern Physics}, vol.~86, no.~2, p.~615, 2014.

\bibitem{doman2010wingbeat}
D.~B. Doman, M.~W. Oppenheimer, and D.~O. Sigthorsson, ``Wingbeat shape
  modulation for flapping-wing micro-air-vehicle control during hover,'' {\em
  Journal of guidance, control, and dynamics}, vol.~33, no.~3, pp.~724--739,
  2010.

\bibitem{khan2007control}
Z.~A. Khan and S.~K. Agrawal, ``Control of longitudinal flight dynamics of a
  flapping-wing micro air vehicle using time-averaged model and differential
  flatness based controller,'' in {\em 2007 American Control Conference},
  pp.~5284--5289, IEEE, 2007.

\bibitem{he2017adaptive}
W.~He, Z.~Yan, C.~Sun, and Y.~Chen, ``Adaptive neural network control of a
  flapping wing micro aerial vehicle with disturbance observer,'' {\em IEEE
  transactions on cybernetics}, vol.~47, no.~10, pp.~3452--3465, 2017.

\bibitem{lee2018learning}
J.~Lee, S.~Ryu, T.~Kim, W.~Kim, and H.~J. Kim, ``Learning-based path tracking
  control of a flapping-wing micro air vehicle,'' in {\em 2018 IEEE/RSJ
  International Conference on Intelligent Robots and Systems (IROS)},
  pp.~7096--7102, IEEE, 2018.

\bibitem{sridhar2019beneficial}
M.~Sridhar, C.-K. Kang, and D.~B. Landrum, ``Beneficial effect of the coupled
  wing-body dynamics on power consumption in butterflies,'' in {\em AIAA
  Scitech 2019 Forum}, p.~0566, 2019.

\bibitem{jayakumar2018control}
J.~Jayakumar, K.~Senda, and N.~Yokoyama, ``Control of pitch attitude by abdomen
  during forward flight of two-dimensional butterfly,'' {\em Journal of
  Aircraft}, vol.~55, no.~6, pp.~2327--2337, 2018.

\bibitem{dyhr2013flexible}
J.~P. Dyhr, K.~A. Morgansen, T.~L. Daniel, and N.~J. Cowan, ``Flexible
  strategies for flight control: an active role for the abdomen,'' {\em Journal
  of Experimental Biology}, vol.~216, no.~9, pp.~1523--1536, 2013.

\bibitem{sridhar2020geometric}
M.~Sridhar, C.-K. Kang, and T.~Lee, ``Geometric formulation for the dynamics of
  monarch butterfly with the effects of abdomen undulation,'' in {\em AIAA
  Scitech 2020 Forum}, p.~1962, 2020.

\bibitem{berman2007energy}
G.~J. Berman and Z.~J. Wang, ``Energy-minimizing kinematics in hovering insect
  flight,'' {\em Journal of Fluid Mechanics}, vol.~582, pp.~153--168, 2007.

\bibitem{ellington1984aerodynamics}
C.~P. Ellington, ``The aerodynamics of hovering insect flight. i. the
  quasi-steady analysis,'' {\em Philosophical Transactions of the Royal Society
  of London. B, Biological Sciences}, vol.~305, no.~1122, pp.~1--15, 1984.

\bibitem{lee2017global}
T.~Lee, M.~Leok, and N.~McClamroch, {\em Global Formulation of {L}agrangian and
  {H}amiltonian Dynamics on Manifolds}.
\newblock Springer, 2018.

\bibitem{dickinson1999wing}
M.~H. Dickinson, F.-O. Lehmann, and S.~P. Sane, ``Wing rotation and the
  aerodynamic basis of insect flight,'' {\em Science}, vol.~284, no.~5422,
  pp.~1954--1960, 1999.

\bibitem{sane2001control}
S.~P. Sane and M.~H. Dickinson, ``The control of flight force by a flapping
  wing: lift and drag production,'' {\em Journal of experimental biology},
  vol.~204, no.~15, pp.~2607--2626, 2001.

\bibitem{teschl2012ordinary}
G.~Teschl, {\em Ordinary differential equations and dynamical systems},
  vol.~140.
\newblock American Mathematical Soc., 2012.

\end{thebibliography}

\end{document}